\documentclass[a4paper,11pt]{article}
\pdfoutput=1
\usepackage{jheppub}
\usepackage[T1]{fontenc}
\usepackage{graphicx}
\usepackage{amsmath}
\usepackage{xspace}
\usepackage{subfig}
\usepackage[table]{xcolor}
\usepackage{mathtools}
\usepackage{braket}
\usepackage{multirow}
\usepackage{slashed}
\usepackage{orcidlink}
\usepackage{appendix}
\usepackage{tikz-3dplot}
\usetikzlibrary{decorations.markings, arrows.meta}

\title{No Track left behind: Graph-based Vertexing for long-lived Particle Reconstruction}

\author[a]{Jonathan Kriewald \orcidlink{0000-0002-3313-115X}}
\emailAdd{jonathan.kriewald@ijs.si}

\affiliation[a]{\normalsize \it  Jo\v zef Stefan Institute, Jamova 39, 1000 Ljubljana, Slovenia}

\abstract{
Reconstruction of displaced vertices is a cornerstone of both precision flavour physics and searches for long-lived particles (LLPs) at colliders. 
While existing vertexing algorithms are highly optimised for primary and short-lived secondary vertices, they face limitations when confronted with the large displacements and heterogeneous topologies characteristic of LLP decays. 
In this work we present a new approach to displaced vertex reconstruction combining a graph-based track clustering strategy as a vertex finder with the established robust vertex fitting procedure. 
The algorithm is implemented as a self-contained \textsc{Delphes} module and can be straightforwardly integrated into existing detector cards, providing a turn-key tool for phenomenological studies. 
This plug-and-play functionality fills an important gap in public fast-simulation frameworks by providing automated pattern recognition for displaced vertex finding, while remaining readily usable in phenomenological studies.
We validate our approach in an IDEA-like FCC-ee detector, using Higgs-strahlung $e^+e^- \to Zh$ with exotic $h\to NN$ decays as a benchmark process. 
We demonstrate excellent efficiency, resolution, and purity across a broad range of lifetimes, and derive model-independent projections for the FCC-ee sensitivity to exotic Higgs branching fractions. 
}


\makeatletter
\gdef\@fpheader{\phantom{a}}
\makeatother

\begin{document}

\maketitle

%
%
\section{Introduction}

Vertex reconstruction is a central ingredient of collider event reconstruction. 
From the determination of the primary interaction point or primary vertex (PV) to the identification of secondary vertices from heavy-flavour hadrons, and the reconstruction of complex multi-prong topologies, vertexing underpins precision measurements as well as searches for new physics. 
In particular, displaced vertices (DVs) from the decays of long-lived particles (LLPs) have emerged as a powerful and generic probe of physics beyond the Standard Model (SM)~\cite{1806.07396,1903.04497,Blondel:2022qqo,Knapen:2022afb} (see~\cite{2205.08582,2301.13866,2312.07484,2402.15804,2403.01556,2403.15332,2409.10806,2410.16835,2503.16213,2505.02429} for some recent searches). 

LLPs appear in a wide range of motivated scenarios: they are a natural consequence of mechanisms generating neutrino masses, such as in type-I seesaw frameworks (and their extensions)~\cite{Atre:2009rg,Blondel:2014bra,Antusch:2016vyf,Antusch:2016ejd,Bellagamba:2025xpd,2202.07310,Cai:2017mow,Das:2019fee,Chiang:2019ajm,Abdullahi:2022jlv} or Left–Right Symmetric Models (LRSM)~\cite{Maiezza:2015lza,Nemevsek:2016enw,Urquia-Calderon:2023dkf,Fuks:2025jrn,Liu:2025ldf}; they arise in Hidden Valley and dark-sector constructions~\cite{Strassler:2006im,Strassler:2006ri,Strassler:2006qa,Han:2007ae,Strassler:2008fv}, and in supersymmetric scenarios with weakly coupled mediators~\cite{Chen:1995yu,Dimopoulos:1996vz,Thomas:1998wy,Barbier:2004ez,Lopez-Fogliani:2005vcg,Fan:2011yu,Ghosh:2017yeh}, see also~\cite{1903.04497} and references therein.
Owing to their macroscopic lifetimes, LLPs can decay at distances of tens of microns up to meters from the primary interaction, producing striking signatures of displaced tracks and vertices. 
Such signatures are highly distinctive, with large background rejection potential, and have become a central focus of current and future collider programs~\cite{1903.04497,Blondel:2022qqo,Knapen:2022afb,Antusch:2025lpm,2505.00272,Altmann:2025feg}.

Over the past decades, a rich ecosystem of vertexing algorithms has been developed. 
These include the Billoir and Kalman-based fits~\cite{Billoir:1985nq,Billoir:1992yq,Fruhwirth:2007hz,Strandlie:2010zz,Fruhwirth:2020zbo}, adaptive and iterative multi-vertex finders~\cite{ATLAS:2019jmx}, and specialised tools such as \textsc{LCFIplus} for flavour-tagging at lepton colliders~\cite{Suehara:2015ura}. 
More recently, graph- and machine-learning-based methods have also begun to appear~\cite{ATLAS:2019wqx,CMS:2024xzb,Correia:2024ogc}. 
Public implementations of vertexing exist in frameworks such as \textsc{CMSSW}~\cite{CMSSW}, \textsc{ACTS}~\cite{Ai:2021ghi}, or \textsc{RAVE}~\cite{rave,Waltenberger:2011zz}, but these are either tightly coupled to specific experiments, or are too heavy-weight to deploy for phenomenological studies.
As a result, phenomenological studies of LLPs often resort to truth-level objects or simplified parametric smearing (see e.g.~\cite{Araz:2021akd} for LLP recasting in \textsc{MadAnalysis5}), leaving the impact of realistic vertex reconstruction unexplored.

In this work, we address this gap by presenting a graph-based vertex finder tailored to displaced topologies, combined with a robust vertex fitter. 
Crucially, our implementation is provided as a self-contained \textsc{Delphes}~\cite{deFavereau:2013fsa} module. 
This makes the vertex finding algorithm readily available to the phenomenology community: it can be plugged directly into existing detector cards with minimal effort, enabling fast and realistic LLP analyses in the widely used \textsc{MadGraph}–\textsc{Pythia}–\textsc{Delphes} simulation chain. 
While \textsc{Delphes} already contains a powerful vertex fitting module, the pattern recognition necessary to identify which tracks should be fitted to a common vertex was left to the user.
Our implementation addresses this gap and provides immediate usability and reproducibility for collider studies.
The repository with all additions is available on GitHub~\href{https://github.com/jkriewald/delphes-LLP}{https://github.com/jkriewald/delphes-LLP}.

In addition to introducing a new graph-based vertex finder, we present a review of the vertex fitting problem. By expressing the fit within a unified Gauss-Newton framework, we recover the standard transverse-information formulation while exposing its equivalent Schur-complement representations and underlying geometric interpretation. We hope that this presentation makes the principles of modern vertex fitting more transparent and provides a natural foundation for the timing extension developed in this work.

To demonstrate its performance and physics reach, we validate our module in an IDEA-like FCC-ee detector~\cite{2502.21223}, using Higgs-strahlung $e^+e^-\to Zh$ with exotic $h\to NN$ decays as a benchmark process. 
Among the variety of possible LLP signatures, exotic Higgs decays provide one of the most motivated and experimentally clean avenues. 
In particular, the decay $h\to NN$ into heavy neutral leptons (HNLs) is predicted in several classes of neutrino mass models, including the LRSM~\cite{Maiezza:2015lza,Nemevsek:2016enw,Fuks:2025jrn} and $B\!-\!L$ or scalar singlet extensions of the type-I seesaw~\cite{Graesser:2007yj,Shoemaker:2010fg,2202.07310,Yang:2025jxc}. 
Depending on their mass and mixing, the HNLs can have lifetimes corresponding to displacements ranging from the vertex detector to the outer tracking layers, producing visible leptonic and semi-leptonic final states. 
As such, this channel constitutes both a motivated physics target and a versatile benchmark for displaced vertex reconstruction at lepton colliders, as it provides displaced vertices spanning the inner pixel detector, the outer vertex layers, and the large drift chamber, and thus constitutes a challenging test-bed. 
We show that our approach achieves excellent efficiency, resolution, and purity across a wide range of lifetimes, and we further exploit the reconstructed vertices to project FCC-ee sensitivities to exotic Higgs branching ratios.

The manuscript is organised as follows. 
While the pioneering work of~\cite{Billoir:1985nq,Billoir:1992yq,Fruhwirth:2007hz,Strandlie:2010zz,Fruhwirth:2020zbo} remains authoritative, in Section~\ref{sec:vertex_fitting} we start with a brief review of vertex fitting addressed at the phenomenology community.
In Section~\ref{sec:graph} we introduce our graph-based vertex finding algorithm, while we validate our approach and showcase its capabilities in Section~\ref{sec:validation_and_performance}.
We then introduce a simple approach to the reconstruction of LLP kinematics in~\ref{sec:LLP} and use the methodologies developed in this work for a full analysis of $\mathrm{BR}(h\to NN)$ in Section~\ref{sec:analysis}, before we conclude in Section~\ref{sec:concs}.
Additional information on the track model as well as the \textsc{Delphes} implementation is provided in two appendices.

\section{Robust Vertex Fitting}
\label{sec:vertex_fitting}
In particle physics experiments, one crucial step in the event reconstruction lies in establishing the common point of origin of reconstructed particles.
Starting from the primary interaction point, or ``primary vertex'' (PV), secondary vertices (SV) or displaced vertices (DV) of long-lived particles are reconstructed for jet flavour-tagging and even full decay chains for precision flavour physics can be reconstructed, reaching impressive accuracies.
For ``detector-stable'' charged particles, such as light leptons ($e^\pm$ and $\mu^\pm$), charged pions $\pi^\pm$ and kaons $K^\pm$ as well as (anti-)protons, the full trajectory as they propagate through the inner detector and the magnetic field can be reconstructed as tracks.
The task of Vertex fitting then lies in finding the common origin of a given set of tracks which may or may not be at the primary interaction point.
Building on the pioneering formulations of~\cite{Billoir:1985nq,Billoir:1992yq,Fruhwirth:2007hz,Strandlie:2010zz,Fruhwirth:2020zbo}, which remain the foundation of modern vertex fitting, we here present a streamlined reformulation aimed at making the underlying principles more accessible and at extending their applicability to long-lived particle reconstruction in fast-simulation studies.

As a first example, let us consider the precision-weighted distance between a single track and a given vertex position such as the beamspot.
A track can be parametrised as a parametric differentiable curve $\gamma(s)$, in which $s$ is the curve parameter and the image of the curve $\vec x(s)\in \mathbb{R}^3$ is the particle position along its trajectory.
The precision-weighted distance between the curve and a (fixed) vertex position $\vec v$ can then be written as a quadratic (log-)likelihood surrogate
\begin{equation}
    F(s) = \frac{1}{2}(\vec v - \vec x(s))^T\,W(s)\,(\vec v - \vec x(s))\,,\label{eqn:metricT2V}
\end{equation}
in which the ``information matrix'' $W(s) = (\mathcal R(s) + \mathcal C_x(s))^{-1}$ is the inverse of the transported covariance matrix of the particle position $\vec x(s)$, accounting for the systematic uncertainties of the track reconstruction, and $\mathcal R(s)$ is the covariance of the ``process noise'' due to interaction of the charged particle with the detector material, magnetic field inhomogeneities and other possible systematic effects.
In the present implementation, however, the propagated track covariance already includes these effects, so that effectively $\mathcal R(s)=0$.

Taking $W(s) = {\mathbb I}$ reduces Eq.~\eqref{eqn:metricT2V} to the usual euclidean distance (or rather to half the square of the euclidean distance).
While our results in this section are independent of the track parametrisation, for concreteness we use the helical perigee parametrisation outlined in Appendix~\ref{sec:trackmodel}.
The distance measure in Eq.~\eqref{eqn:metricT2V} can be iteratively minimised with the help of the Gauss-Newton (GN) algorithm.
Optimising with respect to $s$,
by linearising $\vec x(s)$ around a reference point $s_0$ as 
\begin{equation}
    \vec x(s_0, \Delta s) \approx \vec x(s_0) + \Delta s \frac{\partial \vec x(s)}{\partial s}\Big|_{s=s_0}\equiv \vec x_0 + \Delta s \:\vec t_0\,,
\end{equation}
and keeping $W$ fixed at $s_0$, that is approximating $\frac{\partial W(s)}{\partial s}\simeq 0$~\footnote{For a true Newton/Gauss-Newton optimisation, the variation of the metric $W(s)$ should also be taken into account, however for highly anisotropic $W$ this can lead to ``metric drift'' such that the optimisation algorithm ``chases'' regimes of small $W$ rather than minimising the residual distance $\vec v - \vec x(s)$.}, we can find and solve the normal equations so that at iteration $k+1$
\begin{eqnarray}
    s_{k+1} = s_k + \frac{{\vec t}^{\:T}(s_k) \:W(s_k) \:\vec r(s_k)}{{\vec t}^{\:T}(s_k)\: W(s_k)\: \vec t(s_k)}\,,\label{eqn:GNsingle}
\end{eqnarray}
in which we have defined the residual $\vec r(s) = \vec v - \vec x(s)$ and $\vec t(s) = \frac{\partial\vec x(s)}{\partial s}$.
For a ``true'' Newton step one would need to include a term with second derivatives
$\vec r^T\, W \,\vec a$ with $\vec a(s) = \frac{\partial^2 \vec x(s)}{\partial s^2}$ in the denominator, but this can make the system ill-conditioned far from the solution and is very small close to the solution. 
The GN step in Eq.~\eqref{eqn:GNsingle} keeps the minimisation locally convex and usually leads to super-linear convergence.

After convergence to the maximum likelihood estimator (MLE) $s\to s^\star$, we can further define the statistical compatibility of a track to a given vertex as
\begin{equation}
    \chi^2(s^\star) = 2\min_{s}  F(s) = 2 F(s^\star) = (\vec v - \vec x(s^\star))^T\, W(s^\star) (\vec v - \vec x(s^\star))\,.\label{eqn:chi2}
\end{equation}

The vertex position $\vec v$ is usually a priori not precisely known, and so the task of vertex fitting relies in simultaneously estimating the vertex position (and its covariance) together with the track phases $s$.
To this end, one formulates the vertex fit as a non-linear least squares optimisation problem and the corresponding log-likelihood surrogate can be taken as
\begin{equation}
    F(\{s_i\}, \vec v) = \frac{1}{2} \left[\sum_{i}^{N_\mathrm{tracks}}(\vec v - \vec x_i(s_i))^T\,W_i(s)\,(\vec v - \vec x_i(s_i))\right] + \frac{1}{2}(\vec v - \vec v_p)^T\, \tilde W\, (\vec v - \vec v_p)\,,\label{eqn:likelihood}
\end{equation}
in which $\vec v_p,\,\tilde W$ encode prior information, which could be for instance the position and spread of the beam-spot for primary vertex fitting.
Since the objective is quadratic in the vertex coordinates, the vertex could equally well be eliminated analytically from the outset. We nevertheless derive the full block GN system because it naturally exposes both Schur-complement formulations discussed below.
The GN system of a quadratic objective function $\mathcal F(\theta_i) =\frac{1}{2}\sum_i \rho_i(\theta_i)^2$ for the fit variables $\vec\theta$ can be schematically written as
\begin{equation}
    \mathcal H \Delta \vec \theta = -g\,,
\end{equation}
in which $g_i = \frac{\partial\mathcal F}{\partial \theta_i} = \rho_i \frac{\partial\rho_i}{\partial\theta_i}$ is the gradient and $\mathcal H_{ij} = \frac{\partial\rho_i}{\partial\theta_i}\frac{\partial\rho_i}{\partial\theta_j}$ is the (approximate) Hessian.
Compared to the Newton algorithm, the GN Hessian lacks second derivatives which would be necessary for true quadratic convergence within the basin of attraction.
However, as for the one-dimensional case, the second derivatives can make the system ill-conditioned while the GN Hessian at the local linearisation point is guaranteed symmetric and positive definite (SPD), such that the GN system can be efficiently (numerically) solved.
While the combined MLE ties all $s_i$ and the vertex position $\vec v$ together, they are linearly independent variables and so we can separate the system as 
\begin{equation}
    \begin{pmatrix}
        H_{vv} & H_{vs}\\
        H_{sv} & H_{ss}
    \end{pmatrix}
    \begin{pmatrix}
    \Delta\vec v\\\Delta s_i
    \end{pmatrix} 
    = 
    -\begin{pmatrix}
    g_v\\g_s
    \end{pmatrix}\,.\label{eqn:GNsystem}
\end{equation}
At the current linearization point $s_i = s_i^0$ and the current vertex estimate $\vec v = \vec v_0$, the gradient and Hessian terms for the likelihood in Eq.~\eqref{eqn:likelihood} are obtained as
\begin{eqnarray}
    g_v = \left[\sum_i W_i^0\:\vec {r}_i^0\right] + \tilde W(\vec v_0 - \vec v_p)\,,\quad g_s^i = -\vec{t}_i^{\:0,T} W_i^0 \vec{r}_i^{\:0}\,,
\end{eqnarray}
and 
\begin{eqnarray}
    H_{vv} &=& \left[\sum_i W_i^0\right] + \tilde W\,,\quad H_{ss} = \mathrm{diag}(\vec t_1^{\:0,T}W_1^0\:\vec t_1^{\:0},\:\vec t_2^{\:0,T}W_2^0\:\vec t_2^{\:0},\dots,\:\vec t_N^{\:0,T}W_N^0\:\vec t_N^{\:0})\,,\nonumber\\
    H_{vs} &=& H_{sv}^T = - \left(W_1^0\: \vec t_1^{\:0},\: W_2^0\: \vec t_2^{\:0},\: \dots, W_N^0\: \vec t_N^{\:0}\right)\,,
\end{eqnarray}
in which $\vec r_i^0 = \vec v_0 - \vec x_i(s_i^0)$.
Directly solving Eq.~\eqref{eqn:GNsystem} has a computational complexity of $\mathcal O((3+N)^3)$ and it is therefore advisable to separate the steps $\Delta \vec v$ and $\Delta s_i$ via Schur complements.

Eliminating the phase increments via a Schur complement recovers the transverse-information formulation used in the standard vertex-fitting literature~\cite{Billoir:1985nq,Billoir:1992yq,Fruhwirth:2007hz,Strandlie:2010zz,Fruhwirth:2020zbo,Bedeschi:2024uaf,2401.13538}.
We can thus re-arrange the equation system in Eq.~\eqref{eqn:GNsystem} and solve for $\Delta s_i$ as
\begin{equation}
    \Delta s_i = -H_{ss}^{-1}(g_s^i + H_{sv}\: \Delta \vec v)\,,\label{eqn:step_independent}
\end{equation}
which reduces to the same step as in Eq.~\eqref{eqn:GNsingle} for $\Delta \vec v = 0$.

Inserting $\Delta s_i$ back into Eq.~\eqref{eqn:GNsystem}, which amounts to building the Schur complement $S_v = H/H_{ss}$, we get the normal equation for $\Delta v$ as
\begin{equation}
    S_v \Delta \vec v = \left(H_{vv} - H_{vs}\:H_{ss}^{-1}\: H_{sv}\right) \Delta \vec v= - (g_v - H_{vs}H_{ss}^{-1}\:g_s)\,.\label{eqn:vertex_step}
\end{equation}
This can be further expanded into the components of the Hessian.
Firstly, we find (dropping the ``0'' superscripts)
\begin{equation}
    S_v = \tilde W + \sum_i \left[W_i - W_i\frac{\vec t_i \vec t_i^T}{\vec t_i^T W_i \vec t_i}W_i\right] \equiv \tilde W + \sum_i W_i^\perp\,,
\end{equation}
with 
\begin{equation}
    W_i^\perp = W_i - W_i\frac{\vec t_i\vec t_i^{\:T}}{\vec t_i^{\:T} W_i \vec t_i}W_i\,,
\end{equation} which is orthogonal to $\vec t_i$.
We can then insert the remaining components to find
\begin{equation}
    S_v \Delta \vec v = -S_v \vec v_0 +\left[ \tilde W \vec v_p + \sum_i W_i^\perp \vec x_i(s_i^0)\right]\,.
\end{equation}
Expanding Eq.~\eqref{eqn:step_independent}, we further find
\begin{equation}
    \Delta s_i = \frac{1}{\vec t_i^{\:T} W_i \vec t_i} \left[\vec t_i^T W_i (\vec v_0 + \Delta \vec v) - \vec t_i^T W_i \vec x_i(s_i^0)\right]\,.
\end{equation}
Finally, we can insert $\vec v_0 + \Delta \vec v$, resulting in
\begin{equation}
    \Delta s_i =  \frac{\vec t_i^T W_i}{\vec t_i^{\:T} W_i \vec t_i}\left[ \vec v^\star-\vec x_i(s_i^0) \right]\,,
\end{equation}
with 
\begin{equation}
    \vec v^\star = \vec v_0 + \Delta \vec v = \left[\tilde W + \sum_i W_i^\perp\right]^{-1} \left[\tilde W\vec v_p + \sum_j W_j^\perp \vec x_j(s_j^0)\right]\,,
    \label{eqn:phase_update}
\end{equation}
which is identical to the results of~\cite{Bedeschi:2024uaf}, which have been implemented in \textsc{Delphes}.
Furthermore, the computational complexity is reduced to $\mathcal O(N_\text{tracks})$ with respect to naively solving Eq.~\eqref{eqn:GNsystem}.
This derivation therefore recovers the standard transverse-information vertex fit of Refs.~\cite{Billoir:1985nq,Billoir:1992yq,Fruhwirth:2007hz,Strandlie:2010zz,Fruhwirth:2020zbo,Bedeschi:2024uaf,2401.13538}. 
The present formulation shows that this estimator is precisely the vertex-space Schur complement of the full Gauss-Newton system.
The present formulation is therefore not intended as a new statistical estimator, but rather as a unified derivation that exposes the underlying block structure and its equivalent Schur-complement representations.

We note here that taking the Schur complement amounts to profiling the unknown phases $s_i$.
Since the phase parameter describes motion along the track tangent, profiling the phases is equivalent to analytically eliminating the tangential degree of freedom of each track. Consequently, each track contributes only the two transverse directions encoded in $W_i^\perp$ and the degrees of freedom in a $3D$-vertex fit are then
\begin{equation}
    \mathrm{ndf} = 2N_\mathrm{tracks} - 3\,.
\end{equation}
We further observe that the current/initial vertex estimate $\vec v_0$ cancels in the phase update in Eq.~\eqref{eqn:phase_update} and no initial vertex guess is needed to initialise the fit; the vertex at each iteration is obtained from the current track linearisation points.
The algorithm to find the optimal vertex for a given set of tracks then amounts to choosing a suitable initial linearisation point for the track phases $s_i$, compute $\vec v^\star$ from the current phases and update to the next linearisation point $s_i^{k+1} = s_i^{k} + \Delta s_i$ until convergence.
The choice of the initial linearisation point is important for convergence. 
For displaced vertices, we initialise the track phases from the innermost associated tracker hit (or equivalent detector-informed estimate), which provides a good approximation to the true decay point.

\medskip
As an alternative approach to derive the fit, we can form the other (dual) Schur complement and profile the vertex position first, so that
\begin{equation}
    \Delta v = - H_{vv}^{-1}(g_v + H_{vs}\Delta s)
\end{equation}
and insert back to get
\begin{equation}
    (H_{ss} - H_{sv} H_{vv}^{-1}H_{vs})\Delta s = -(g_s - H_{sv}H_{vv}^{-1} g_v)\,.
\end{equation}
This seems impractical at first, since the system that has to be solved is now $N\times N$ with a computational complexity of $\mathcal O(N^3)$.
However, after noticing that $H_{ss}$ is diagonal and the second term $H_{sv} H_{vv}^{-1} H_{vs}$ amounts to a rank-3 update of $H_{ss}$, we can employ the Woodbury matrix identity such that
\begin{equation}
     (H_{ss} - H_{sv} H_{vv}^{-1}H_{vs})^{-1} = H_{ss}^{-1} + H_{ss}^{-1} H_{sv}(H_{vv} - H_{vs}H_{ss}^{-1}H_{sv})^{-1}H_{vs}H_{ss}^{-1}\,.
\end{equation}
Thus, the only matrix solve or inversion is identical to the first case such that the computational complexity is similarly $\mathcal O(N)$ (accumulating the phase blocks) and a $3\times 3$ solve.
Inserting the components of the Hessian, the phase step can be shown to be identical to the one in Eq.~\eqref{eqn:phase_update}.

While algebraically equivalent to the first Schur complement, the derivation via the dual Schur complement provides a complementary geometric interpretation.
We can first minimise the likelihood in Eq.~\eqref{eqn:likelihood} with respect to $\vec v$ exactly (since it is quadratic in $\vec v$).
We thus obtain
\begin{equation}
    \vec v(\{s_i\}) =\tilde S^{-1}\left(\tilde W \vec v_p+ \sum_i W_i\vec x_i(s_i)\right)\,,
\end{equation}
with 
\begin{eqnarray}
     \tilde S = \left(\tilde W + \sum_i W_i\right)\,.
\end{eqnarray}
After inserting  $\vec v$ back into Eq.~\eqref{eqn:likelihood} we find the reduced (profiled) objective
\begin{equation}
    F(\{s_i\}, \vec v^\star(\{s_i\})) = \frac{1}{4}\sum_{i<j}(\vec x_i - \vec x_j)^T \left(W_i \tilde S^{-1}W_j + W_j \tilde S^{-1} W_i\right) (\vec x_i - \vec x_j)\,,
\end{equation}
so that optimising over $s_i$ is equivalent to simultaneously minimising all pair-wise track distances under the combined metric $(W_i \tilde S^{-1}W_j + W_j \tilde S^{-1} W_i)$.
In this derivation the vertex coordinates no longer appear explicitly; instead, they induce an effective interaction between every pair of tracks through the common profiled vertex.
Minimising the reduced objective with respect to $s_i$ leads to coupled steps in $s_i$ with a coupling between the different tracks.
The coupling arises because changing the phase of one track changes the profiled common vertex, thereby modifying the residuals of all remaining tracks.

After convergence, the vertex covariance can be estimated by propagating the track covariances (see e.g.~\cite{Bedeschi:2024uaf})
\begin{equation}
    \mathcal C_{\vec v} = \left(\tilde W + \sum_{i}W_i^\perp\right)^{-1}\left[\sum_i W_i^\perp W_i^{-1} W_i^\perp\right]\left(\tilde W + \sum_{k}W_k^\perp\right)^{-1}\,.
\end{equation}
This ``sandwich correction'' unfortunately amplifies any bad numerical conditioning in $W_i$ and $W_i^\perp$, which becomes significant especially for very anisotropic $W_i$ in far vertices.
Here we employ the numerically more stable Hessian approximation~\cite{Fruhwirth:2020zbo} 
\begin{equation}
    \mathcal C_{\vec v} \approx \left(\tilde W + \sum_{i}W_i^\perp\right)^{-1}\,,
\end{equation}
which becomes exact for $\tilde W\simeq 0$ (or rather for a large prior vertex covariance since $\tilde W = \mathcal C_p^{-1}$), because $W_i^{\perp} W_i^{-1}\: W_i^\perp = W_i^\perp$.
The vertex fit is validated in Section~\ref{sec:validation_and_performance}.

\bigskip
The fitting method outlined in the above typically enjoys a very fast convergence in a few iterations.
However, the set of tracks to be fitted to the vertex can contain outliers, which can be mismeasured tracks, or worse, well-measured but mis-assigned tracks which don't actually belong to the vertex.
These can significantly distort the vertex fit or prevent reaching an acceptable $\chi^2$ altogether and need to be detected and dealt with.
This can be done by iteratively removing the track with the worst residual $\chi^2_t$ contribution and refitting until an acceptable $\chi^2_{\vec v}$ has been reached, via the random sample consensus (RANSAC) algorithm, or by iteratively down-weighting tracks based on their residual $\chi^2_i$.
The latter was first formalised in the Adaptive Vertex Fitter~\cite{Fruhwirth:2007hz,Waltenberger:2008zz}, which is widely employed by experimental groups.
The idea is to re-formulate the vertex fit from a non-linear least-squares problem to an iteratively re-weighted least-squares (IRLS) problem.
After a vertex fit has converged to its MLE, we re-weight tracks based on their residual contributions to the likelihood objective with a robust weight function (e.g. Huber, Tukey, Sigmoid or more generally a re-descending $M$-estimator).
Each track's $\chi^2$ can be evaluated with Eq.~\eqref{eqn:chi2}, and we assign (Sigmoid) weights as
\begin{equation}
    w_i = \psi(\chi_i^2) = \frac{1}{1 + \exp(\frac{\chi^2_i - \chi^2_c}{\beta})}\,,
\end{equation}
in which $\chi^2_c$ controls the midpoint of the Sigmoid and $\beta$ its slope.
Tracks with $\chi^2_i \leq \chi^2_c$ have weights $w_i\in [0.5, 1]$, while tracks with $\chi^2_i > \chi^2_c$ are smoothly down-weighted to $w_i\to 0$.
The parameter $\chi^2_c$ can be chosen to ``gate'' tracks with a contribution of more than $3\sigma$ (then $\chi^2_c\simeq 9$) and $\beta$ can either be fixed, varied according to some (annealing) schedule~\cite{Fruhwirth:2007hz}, or self-tuning to data, e.g. with a robust scale such as the median absolute deviation.
The weights then appear in the IRLS vertex fit likelihood as
\begin{equation}
    F_{w}(\{s_i\}, \vec v) = \frac{1}{2} \left[\sum_{i}^{N_\mathrm{tracks}} w_i(\vec v - \vec x_i(s_i))^T\,W_i(s)\,(\vec v - \vec x_i(s_i))\right] \,.\label{eqn:weighted_likelihood}
\end{equation}
Including the weights in the fit iterations effectively amounts to re-scaling each track's information $W_i\to \hat W_i = w_i W_i$, and the same formulas derived in this section can be applied.
One then typically starts with unit weights for each track, solves the inner problem, re-weights, and iterates the algorithm until the weights are stable.
This method has the advantage that outliers are smoothly down-weighted without the order-dependence of iteratively removing the worst track and re-fitting, and without the computational cost of RANSAC approaches.

After the weighted fit has converged, meaning that the weights have stabilised, we can then classify tracks as inliers if their weights $w_i \geq 0.5$, while tracks with $w_i\leq 0.5$ are outliers.
The reported $\chi^2$ and degrees of freedom of the vertex fit should then only rely on inliers, while the contribution of outliers is discarded.

\subsection{From $3D$ to $4D$: fitting the vertex time}
In very busy tracking environments such as the high pile-up conditions at the high-luminosity phase of the LHC or the immense beam-induced-backgrounds at future high-energy lepton colliders, timing of tracks (or even individual hits) becomes a crucial discriminator in order to separate signal from background~\cite{CERN-LHCC-2020-007,CMS:2667167,Casarsa:2024mre,MAIA:2025hzm}.
Furthermore, timing can be leveraged for particle identification by measuring the time-of-flight of the reconstructed particle~\cite{2502.21223} and can even be used for long-lived particle searches~\cite{Liu:2018wte,Mason:2019okp,Chiu:2021sgs,Liu:2022ugx,Blondel:2022qqo}, since their decay vertices can have a significant time delay of several nano-seconds, timing requirements allow for very powerful background rejection.

In principle, we can promote the $3D$ vertex fits outlined in the above to full $4D$ fits relying on the individual track timing.
However, the envisaged timing resolution of $\mathcal O(10-100)$~ps (or rather $\mathcal O(3-30)$~mm in natural units) is orders of magnitude worse compared to the positional resolution of $\mathcal O(3-100)\:\mathrm{\mu m}$.
A full inclusion of timing thus buys very little additional information on the vertex position while one has to pay a steep price of numerically ill-conditioned cross-covariance matrices.
Therefore, it is advisable and in general a good approximation to decouple the spatial and timing components from the vertex fit.
In this decoupling limit, we can add a timing term to the likelihood objective as
\begin{equation}
    F_\tau(\{s_i\},\tau_v) = \frac{1}{2}\sum_i\frac{(\tau_v - \tau_i(s_i))^2}{(\sigma^i_{\tau})^2}\,,
\end{equation}
in which $\sigma_{\tau}^i$ is the individual track timing uncertainty, $\tau_v$ the vertex time to be fitted and $\tau_i(s_i)$ the track time evaluated at the track phase $s_i$.
A simple timing model can be constructed as
\begin{equation}
    \tau_i(s_i) = \tau_i^\mathrm{ref} - \frac{s_i^\mathrm{ref} - s_i}{\beta_i}\,,
\end{equation}
in which $\tau_i^\mathrm{ref}$ is the measured track time at a reference phase $s_i^\mathrm{ref}$ and $\beta_i = \frac{|\vec p_i|}{E_i}$ the track velocity.
Since we decouple the spatial and time parts of the vertex fit, the vertex time can be taken as the minimiser of the profiled objective $F_\tau(\{s_i^\star\}, \tau_v)$ and results in the precision-weighted average
\begin{equation}
    \tau_v^\star = \frac{1}{\sum_i \frac{1}{(\sigma_\tau^i)^2}}\sum_j \frac{1}{(\sigma_\tau^j)^2}\tau_j(s_j^\star)\,,
\end{equation}
where the $s_i^\star$ are taken from the spatial MLE.

\subsection{Track parameter re-fitting}
In the classic ``full Billoir fit''~\cite{Billoir:1985nq,Billoir:1992yq}, the vertex position is fitted together with a track parameter refit at each iteration, forcing the tracks to go through the vertex constraint (see also ``steered vertex fit'' in~\cite{Bedeschi:2024uaf}).
In here, and especially due to the possibility of the presence of outliers in the vertex fit, we refit track parameters to the vertex only after the vertex fit has converged and outliers are sorted out.
This can be done by treating the fitted vertex (and its covariance) as a pseudo-measurement.
Employing the Kalman filter formalism for track fitting~\cite{Billoir:1983mz,Fruhwirth:1987fm}, we update the track parameters with a single Kalman-like update as follows.
We linearize the track model at the vertex as
\begin{equation}
    \vec x_i(s_i^\star, \alpha_i) \simeq \vec x_i(s_i^\star, \alpha_i^0) + J_\alpha(s_i^\star\,,\alpha_i^0)\,\delta\alpha_i\,,
\end{equation}
in which $\alpha_i^0$ are the original track parameters (see Appendix~\ref{sec:trackmodel} for the helix model) and $J_\alpha = \frac{\partial \vec x}{\partial \alpha}$ is the Jacobian of the track model.
We further define the measurement residual as $\vec y = \vec v - \vec x(s^\star, \alpha^0)$ and the cross-covariance $\mathcal C_{\alpha,\vec x}$ and innovation covariance  $\mathcal S$ as
\begin{equation}
    \mathcal C_{\alpha,\vec x} = \mathcal C_{\alpha^0} J_\alpha^T\,,\quad \mathcal S = J_\alpha \mathcal C_{\alpha^0} J_\alpha^T + \mathcal C_{\vec v}\,,
\end{equation}
resulting in the Kalman gain matrix
\begin{equation}
    \mathcal K = \mathcal C_{\alpha,\vec x}\:\mathcal S^{-1}\,.
\end{equation}
The track state and its covariance (in the Joseph form) can then be updated as
\begin{equation}
    \alpha_i^0 \to \alpha_i^0 + \mathcal K\vec y\,,\quad \mathcal C_\alpha = (\mathbb{I} - \mathcal K J_\alpha)\:\mathcal C_{\alpha^0}\:(\mathbb{I} - \mathcal K J_\alpha)^T + \mathcal K\:\mathcal C_{\vec v}\:\mathcal K^T\,.
\end{equation}
We note here that in this approach we neglect the cross-correlation $C_{ij} = \langle\delta\vec\alpha_i\vec\alpha_j^T\rangle$ between the track parameters of track $i$ and track $j$ going through the same vertex, so the updated covariance $\mathcal C_\alpha$ is an approximation.
A more complete treatment becomes necessary if one wants to accurately estimate the uncertainties of vertex properties such as the invariant mass or its total momentum. 
\section{Graph-based Vertex Finding}
\label{sec:graph}
Before we can fit tracks to a vertex, a suitable set of tracks likely belonging to the same vertex has to be identified.
Many different methods have been proposed and are being used in the literature, which are tuned to the specific vertexing problem of either primary vertexing, close secondaries for jet tagging, or even full kinematic decay chain reconstructions.
The widely used methods include (gaussian) track density methods for primary vertexing~\cite{Piacquadio:2008zzb}, the iterative vertex finder and the adaptive multi-vertex finder~\cite{ATLAS:2019jmx} (see also~\cite{Ai:2021ghi} for their public implementation), the LCFIplus algorithm~\cite{Suehara:2015ura}, and many others (see e.g.~\cite{Strandlie:2010zz,Fruhwirth:2020zbo} for exhaustive reviews).
More recently, graph-~\cite{ATLAS:2019wqx} and graph-neural-network~\cite{CMS:2024xzb,Correia:2024ogc} based methods start to be applied, dedicated to beyond the SM long-lived particle searches.
Here, we propose a graph-based (not graph neural network) approach, also dedicated to long-lived particle searches.

We start by roughly categorising tracks as prompt/displaced based on their correlated impact parameter significance
\begin{equation}
    \mathcal S_\mathrm{IP} = \sqrt{(D_0, Z_0)(\mathrm{Cov}(D_0,Z_0))^{-1}(D_0, Z_0)^T}\,.\label{eqn:IPsig}
\end{equation}
If $\mathcal S_\mathrm{IP}$ is below a certain threshold, we have found $\mathcal S_\mathrm{IP}\leq 3$ to work well, we categorise the track as prompt.
With all prompt tracks we then fit the primary vertex with a beamspot prior using the vertex fitter described in the previous section.
The resulting PV position and covariance are then used to test all remaining tracks (not attached to the PV fit) against the PV by minimising the single track to vertex distance under the precision metric, i.e. we minimise Eq.~\eqref{eqn:metricT2V} and evaluate the $\chi^2$ at the solution. 
If a track's $\chi^2$ to the PV is above a certain threshold, e.g. $\chi^2_{i\,,\mathrm{PV}}\geq 9$, the track is categorised as displaced, and otherwise as prompt.
The displaced tracks are then used for displaced vertex finding as outlined in the following.

\medskip
With all tracks that have been categorised as displaced, we build a scored undirected compatibility graph in which each track represents a node and edges between nodes are built if a two-track fit yields a $\chi^2_\mathrm{pair} \leq 9$.
Additionally, if track timing information is available, we require timing compatibility of the potential edge such that an edge is only retained if
\begin{eqnarray}
    \chi^2_\mathrm{pair,time}  = \frac{(\tau_1(s_1^\star) - \tau_2(s_2^\star))^2}{(\sigma_\tau^1 + \sigma_\tau^2)^2} \leq 9\,,
\end{eqnarray}
which reduces the risk of reconstructing fake vertices from random track crossings. 
Furthermore one can employ material vetos in the sense that edges are rejected if the fitted vertex position is very close to a detector layer and thus likely stems from material interactions and not from a real particle decay.
Furthermore, a fitted vertex must always lie upstream of the tracks' first hit positions in the tracker so that vertex candidates that lie downstream of either track's first hit position can be discarded.

Since the computational complexity of the pair graph naively scales as $\mathcal O(N^2)$ two-track fits, before a fit is attempted, we test if two tracks are a candidate for a vertex by checking if their trajectories in the $(x,y)$-plane (with the beam-line and magnetic field in the $z$-direction) have an overlap. 
In a homogenous magnetic field, the particle's trajectories in the $(x,y)$-plane are circles so a necessary condition for a vertex can be determined as
\begin{equation}
    |R_1 - R_2| < d < R_1 + R_2\,,
\end{equation}
in which $R_{1,2}$ are the circle's radii and $d$ the distance between their centres.
Track pairs that pass this necessary condition are then fitted to a common vertex and the final $\chi^2_\mathrm{pair}$ is retained as part of the edge score.
In very sparse environments, like leptonic LLP decays (e.g. heavy neutral leptons with $N\to\ell^+\ell^-\nu$ or $Z'\to\ell^+\ell^-$), we can essentially stop here and report the fitted vertex, together with its tracks, as a LLP candidate for further kinematic reconstruction.

In very busy environments like events with hadronic or semi-leptonic LLP decays (potentially with additional prompt jets), further processing is necessary before attempting a full multi-track vertex fit on the vertex candidate.
Since the resulting potentially very dense pair graph likely contains spurious edges between nearby displaced vertices (e.g. heavy flavour chains in hadronic LLP decays or collimated pairs of LLPs in boosted environments), a number of pruning techniques are applied starting by applying so-called mutual $k$-nearest-neighbour ($k$NN) pruning~\cite{kNNpaper}.
This means that only edges are retained if they are within their mutual $k$NN lists (with e.g. $k=5$) in which the distance measure is taken as $\chi^2_\mathrm{pair}$.

In a second stage, we establish a second edge score which we call edge-support. 
For each edge $E(i,j)$ retained after the $k$NN pruning, e.g. an edge between tracks $A$ and $B$, we demand that there be (at least) a third $C$ such $E(A,B)$, $E(A,C)$ and $E(B,C)$ form valid edges and if so, perform an unweighted triplet fit of all three tracks. 
If $\chi^2_\mathrm{triplet}/\mathrm{ndf}\geq 9$, the edge $E(A,B)$ is erased and otherwise retained and receives the support score $s_\mathrm{support} = 1$. 
Each additional track that can be added in this scheme and yields an acceptable (unweighted) triplet fit increases the support score by 1.
Demanding a minimum support score, e.g. $s_\mathrm{support}\geq 1$, reduces the pair graph to strongly connected components and significantly helps in disentangling close-by or overlapping vertices.

As an additional safety measure, we employ Tarjan's bridge finding algorithm~\cite{TARJAN} which finds edges that build connections between strongly intra-connected sub-graphs.
Such edges are only retained if they have both a small $\chi^2_\mathrm{bridge} \leq 3$ and an increased minimum support $s_\mathrm{support}\geq 2$, in order to not accidentally over-fragment the graph but disconnect spurious edges in order to disentangle and resolve close-by decay vertices.
Finally, with the remaining edges, we extract connected components with a simple ``depth-first-search'' (DFS) algorithm.

Before attempting multi-track vertex fits with the tracks contained in the extracted components, we test each of the components for multi-modality with a simple heuristic as follows.
In a maximally connected component of $N$ tracks, we should have $N_\mathrm{3\,max} = {N\choose 3} = \frac{N(N-1)(N-2)}{6}$ triplets that have an acceptable fit. 
If the component has more than one centroid and is actually composed of two strongly connected sub-components, this number should be significantly lower.
We define the triplet score as
\begin{equation}
    \rho_3 = \frac{N_\mathrm{3}}{N_\mathrm{3\,max}}\,.
\end{equation}
If now an extracted component has e.g. $\rho_3\leq 0.8$ (or some other threshold), we attempt to split the component, by first raising the minimum required support score by 1 and then again extract the connected components of this sub-graph via DFS.
As a safety measure against over-fragmentation, we test the union of the connected components of the sub-graph again with the proposed triplet-score and if the union has $\rho_3 \geq 0.8$, we re-connect them.
Tracks not contained in either component are detached, as are all ``singletons''.
With the steps outlined in the above, track assignments to vertex candidates are unambiguous from the start so that we can fit each resulting cluster independently with the vertex fitter described in Section~\ref{sec:vertex_fitting}.
Initialising the the track phases at the tracks' points of closed approach to the precision-weighted average of all triplet fit positions leads to a very fast convergence in usually $2-3$ iterations.
Multitrack vertices that have a final $\chi^2_\mathrm{MT}/\mathrm{ndf}\leq 9$ are retained as LLP candidates and further processed with kinematic reconstruction as outlined in Section~\ref{sec:LLP}.
\section{Vertexing Performance at FCC-ee}
\label{sec:validation_and_performance}
We have implemented both the new fitting algorithm and the graph-based vertex finder as a self-contained \textsc{Delphes}~\cite{deFavereau:2013fsa} module dubbed \textsc{GraphDisplacedVertexFinder} (available on GitHub at~\cite{Graph}).  
While the module is detector agnostic, for concreteness we evaluate it in an IDEA-like FCC-ee detector~\cite{2502.21223}, extending the official IDEA card to include our vertexing module.  
The IDEA detector concept combines a high-granularity silicon vertex detector with a large cylindrical drift chamber, providing continuous tracking out to radii of about two metres~\cite{2502.21223}.  
Originally optimised for jet momentum resolution, this large, low-mass tracking volume also makes the detector exceptionally sensitive to displaced vertices from long-lived particles~\cite{Blondel:2022qqo}.

As a benchmark we simulate Higgs-strahlung production,
$e^{+}e^{-}\to Z h$, with the exotic decay $h\to N N$, where the heavy neutral leptons $N$ undergo semi-leptonic decays $N\to\mu^\pm q\bar q'$ and can travel macroscopic distances before decaying.
This channel provides a clean source of displaced vertices over a broad range of transverse flight distances, and serves as a non-trivial example beyond processes like $Z\to N\nu$, in which only a single decay vertex is expected.

In the following we validate the vertexing algorithm on dedicated samples with proper lifetimes
$c\tau = 10,\;100,\;\text{and}\;1000\;\mathrm{mm}$,
chosen to populate the inner pixel layers, the outer vertex detector, and the large drift chamber respectively.  
These samples allow us to probe the spatial resolution and efficiency across the different tracking regions of the detector, and to test the agreement between the predicted per-vertex covariance and the observed residuals between reconstructed vertices and the simulated ``Monte-Carlo truth''.
The samples are simulated relying on the Feynrules~\cite{Christensen:2008py,Alloul:2013bka}/UFO~\cite{Degrande:2011ua,Darme:2023jdn} model files of the minimal left-right symmetric model (LRSM) developed in~\cite{2403.07756}.
Hard scattering events are simulated with \textsc{MadGraph5\_aMC\_v3.5.3}~\cite{Alwall:2014hca}, are then passed to~\textsc{Pythia\_8.3}~\cite{Bierlich:2022pfr} for parton showers and hadronisation, before they are processed with our modified~\textsc{Delphes}~\cite{deFavereau:2013fsa} for fast detector simulation.
In particular, we rely on the \textsc{TrackCovariance} and \textsc{TimeSmearing} modules developed in~\cite{Bedeschi:2022rnj}, for a fast simulation of track reconstruction.
In the \textsc{TrackCovariance} module, the track parameters and covariances are estimated from the parametrised single hit resolutions in the inner tracker and the Drift Chamber (DCH), taking $1D$ and $2D$ measurements to mimic the detector response, while track finding (in the sense of pattern recognition) is done at truth level.
Multiple scattering effects in the active and passive detector material layers are taken into account as additional contributions to each ``hit'' covariance and then consequently propagated to the track parameter covariances, all under the simplifying assumption of a perfectly homogeneous magnetic solenoid field in the $z$ direction (parallel to the beam-pipe).
Track parameters and their covariances (in the helix model) are then estimated with the Kalman-filter approach~\cite{Billoir:1983mz,Fruhwirth:1987fm}.

Based on the reconstructed tracks, we run our vertex finding and fitting module with the following settings (see also Appendix~\ref{sec:delphes} and the example card~\cite{Card}).
We demand that each reconstructed vertex contains at least three tracks out of which at least one is identified as a lepton ($e$ or $\mu$).
The minimum support score for a graph edge (see Section~\ref{sec:graph}) is taken to be $1$, leading to dense vertex candidates that are passed to the fitter.
These requirements heavily suppress possible fakes from $K_S\,,\Lambda_0$ and charm-meson decays.
While we focus here on semi-leptonic HNL decays, we want to emphasise that the implementations of the vertex finder and fitter are agnostic in what regards final states and that leptonically or hadronically decaying LLPs can be reconstructed as well with the appropriate settings for the graph building stage (see Table~\ref{tab:GraphDV} for an overview).

\begin{figure}
    \centering
    \includegraphics[width=0.48\linewidth]{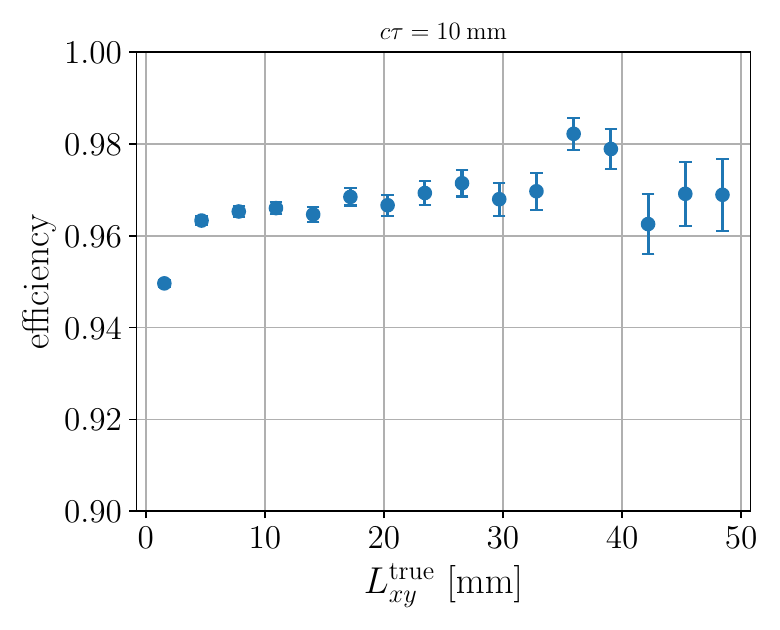}
    \includegraphics[width=0.48\linewidth]{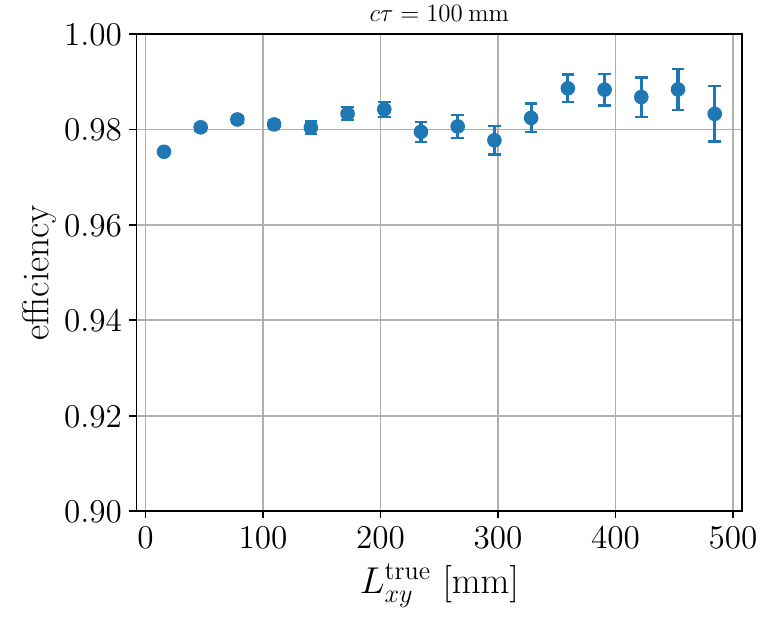}
    \includegraphics[width=0.48\linewidth]{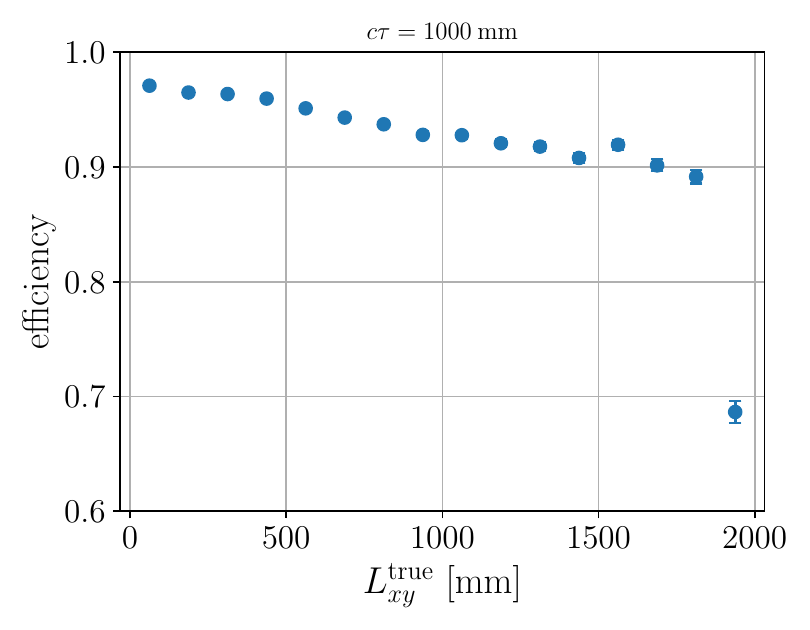}
    \caption{Vertex reconstruction efficiencies of $N$ decays binned in the transverse displacement $L_{xy}$ of the true $N$ decay position. For the different panels we have assumed proper $N$ lifetimes $c\tau(N)\in[10,100,1000]\:\mathrm{mm}$ (as indicated).}
    \label{fig:efficiency_Lxy}
\end{figure}

\begin{figure}
    \centering
    \includegraphics[width=0.48\linewidth]{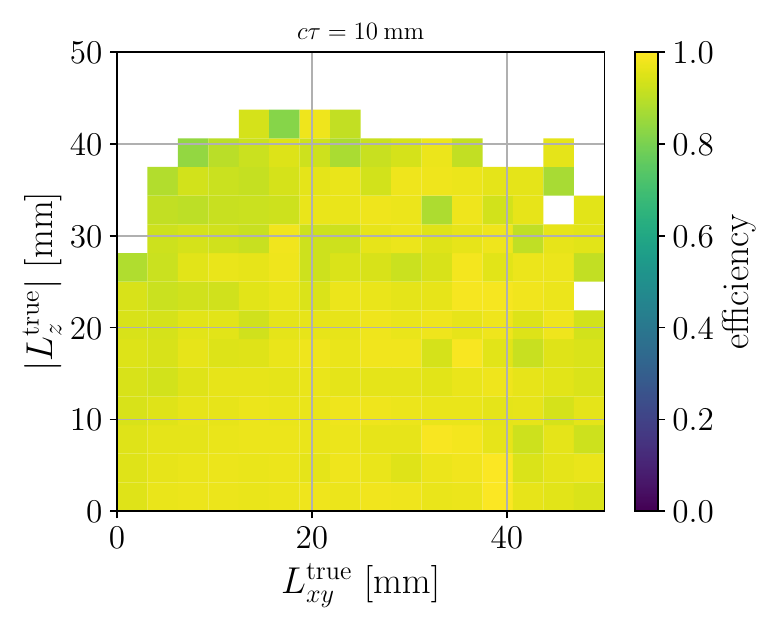}
    \includegraphics[width=0.48\linewidth]{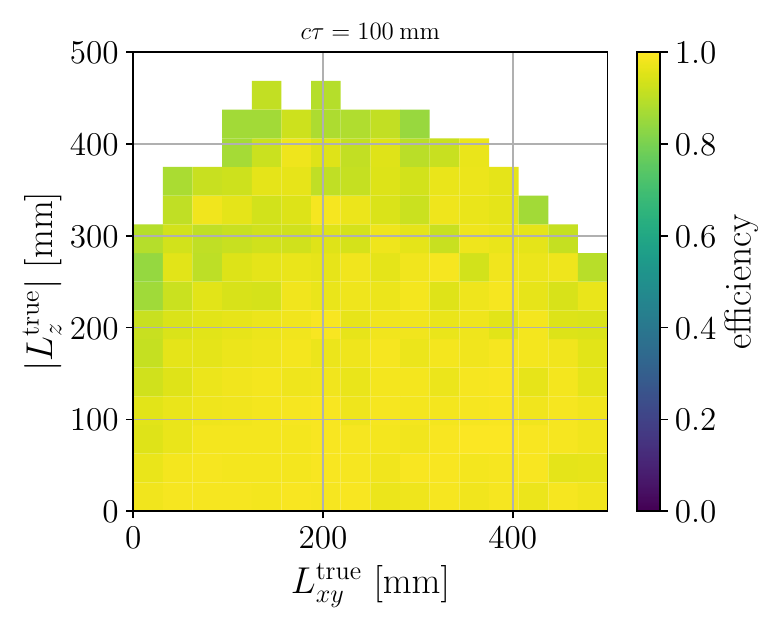}
    \includegraphics[width=0.48\linewidth]{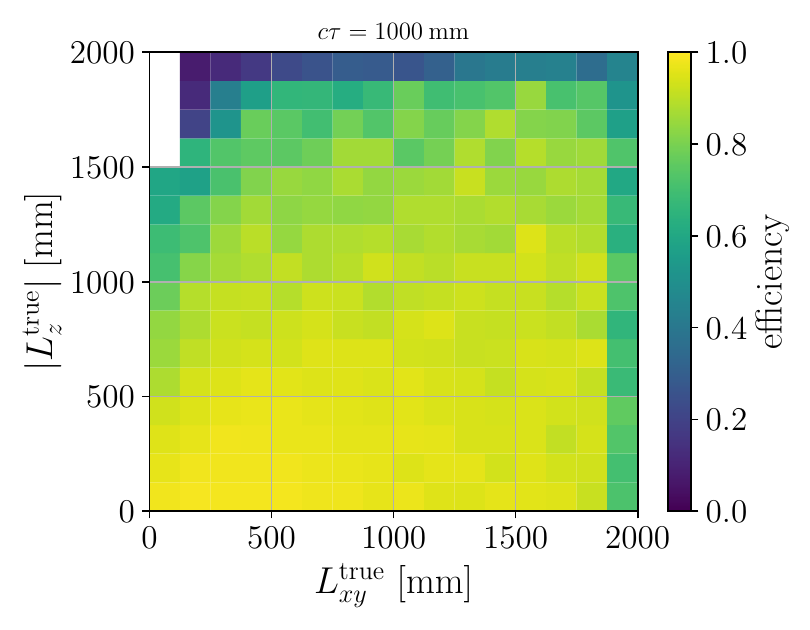}
    \caption{Vertex reconstruction efficiencies of $N$ decays binned in the transverse and longitudinal displacement $L_{xy}$ and $L_z$ of the true $N$ decay position. For the different panels we have assumed proper $N$ lifetimes $c\tau(N)\in[10,100,1000]\:\mathrm{mm}$ (as indicated).}
    \label{fig:efficiency_Lxy_Lz}
\end{figure}

We simulate the benchmark process $e^+ e^- \to Zh, h\to NN$ with subsequent $N\to\mu^\pm q\bar q'$ decays at $\sqrt{s} = 240\:\mathrm{GeV}$.
The $Z$ is forced to decay to $e,\,\mu$ or $\nu$ pairs to keep a clean sample.
We set $m_N = 45\:\mathrm{GeV}$ and simulate $10^5$ events for each of the proper lifetimes $c\tau\in[10,100,1000]\:\mathrm{mm}$.
In order to assess the algorithmic performance of the vertex finding and fitting proposed in this work, we first present the reconstruction efficiency of displaced vertices.
We define the per-event efficiency as 
\begin{equation}
    \mathrm{efficiency} = \frac{\# \mathrm{reconstructed\:vertices}}{\# \mathrm{reconstructable\:vertices}}\,,
\end{equation}
in which ``reconstructable vertices'' are taken at truth level as $N$'s that decay within the fiducial tracking volume and that produce at least two tracks with a minimum $p_T \geq 100\:\mathrm{MeV}$ and a maximum pseudo-rapidity of $|\eta|\leq 2.56$.
First results on the efficiencies are presented Figure~\ref{fig:efficiency_Lxy}, in which we show the efficiencies binned in the transverse displacement $L_{xy}$ of the true $N$ decay position, together with binomial error bars accounting for statistical uncertainties.
As one can see, the efficiencies are very high reaching more than $98\%$ for transverse displacements $L_{xy}\gtrsim 50\:\mathrm{mm}$ while for smaller displacements they range from $95\%$ to $97\%$.
This can be understood, since the momentum distributions of the final state particles introduce correlations between the transverse displacements, the transverse momentum $p_T$ and the pseudo-rapidity $\eta$, such that regimes of very small displacement correspond likely to a large $|\eta|$ in which the reconstruction of tracks happens at the edge of the tracking coverage ($|\eta|\leq 2.56$).
For very large displacements $L_{xy} \geq 500\:\mathrm{mm}$ the efficiency further diminishes to $90\%$, while the last bin has a reduced efficiency of about $\sim 70\%$, since charged particles that originate close to the edge of the fiducial tracking volume simply do not produce enough hits in the DCH to be reconstructed.
In Figure~\ref{fig:efficiency_Lxy_Lz} we present the results for all three samples binned in the transverse and longitudinal displacements $L_{xy}$ and $L_z$, showing very high vertex reconstruction efficiencies throughout the vast majority of the fiducial tracking volume.

\begin{figure}
    \centering
    \includegraphics[width=0.48\linewidth]{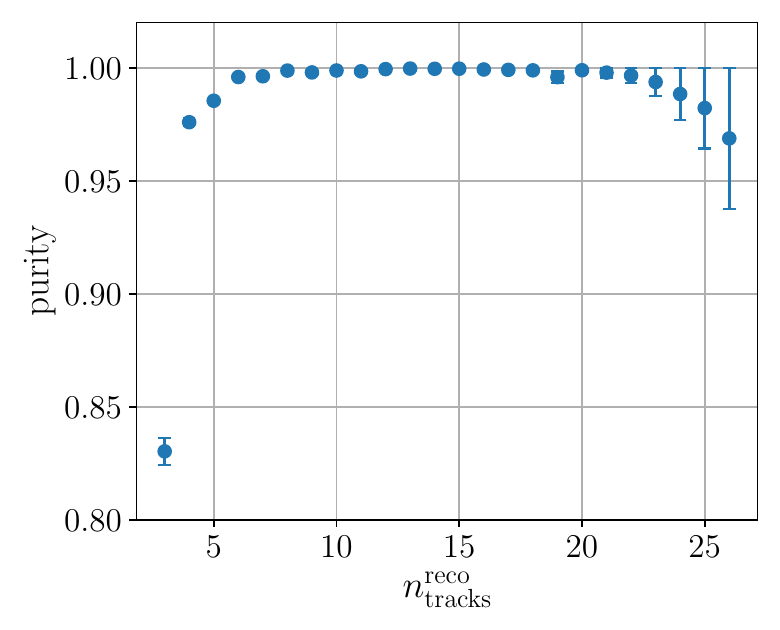}
    \includegraphics[width=0.48\linewidth]{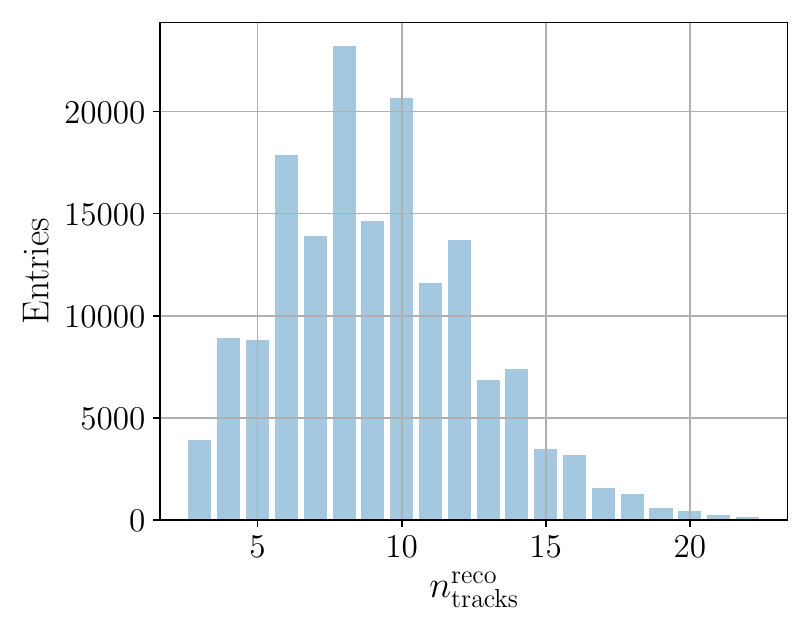}
    \caption{Left: Purity of the reconstructed vertices depending on the amount of tracks per reconstructed vertex. Right: Distribution of track multiplicities of the reconstructed vertices.}
    \label{fig:purity}
\end{figure}

A further point concerns the purity of the reconstructed vertices. 
While track multiplicities strongly vary, we assess purity more heuristically as number of reconstructed vertices to which the closest truth-level vertex corresponds to the generated $N$ decay divided by all reconstructed vertices.
For all three samples we get essentially identical results (within statistical uncertainties) and we present in Figure~\ref{fig:purity} the vertex purity binned in track multiplicity per reconstructed vertex (for the $c\tau = 1000\:\mathrm{mm}$ sample), together with the distribution of track multiplicities.
As one can see, the purities are very high for vertices with at least $4$ tracks, reaching $\geq 99\%$ for vertices with more than $5$ tracks.
For vertices with only $3$ tracks, the purity is somewhat diminished to $80-85\%$, in which the closest ``true vertex'' often stems from secondary/tertiary heavy flavour decays in $N\to\mu^\pm c\bar s$ (e.g. multi-prong decays of $D$-mesons or $\tau$-leptons).
However, these can be efficiently suppressed by placing a cut on the invariant mass of the reconstructed vertex or on the track multiplicity.
As can be seen in the right panel of Figure~\ref{fig:purity}, the vast majority of reconstructed vertices has a track multiplicity significantly exceeding $3$ in which purities are essentially at $100\%$.

\begin{figure}
    \centering
    \includegraphics[width=0.48\linewidth]{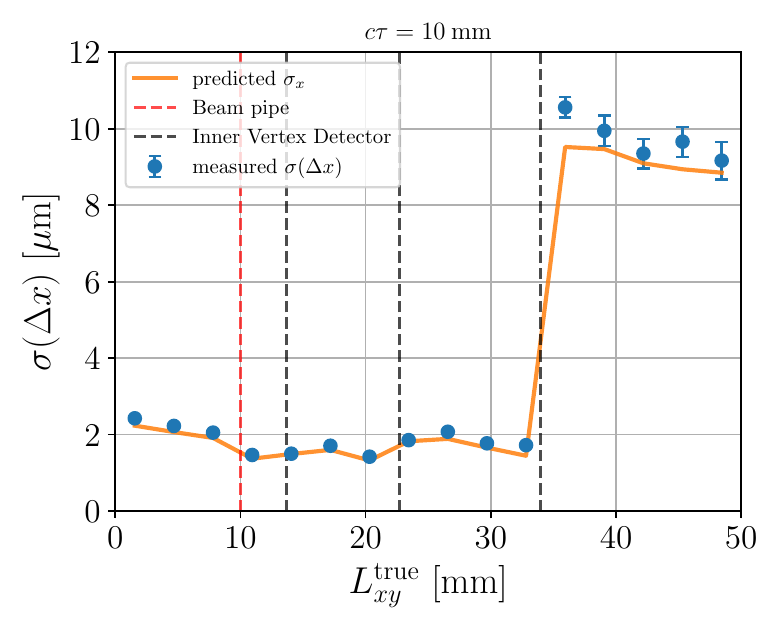}
    \includegraphics[width=0.48\linewidth]{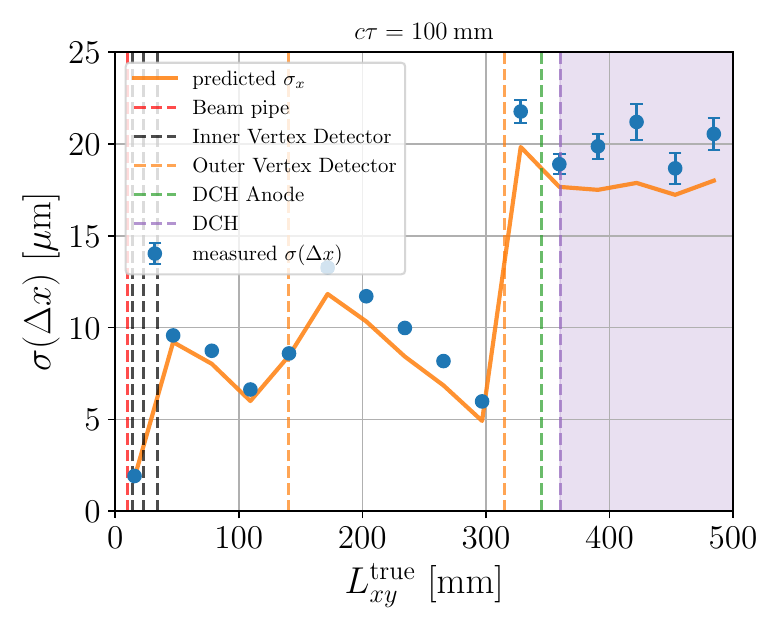}
    \includegraphics[width=0.48\linewidth]{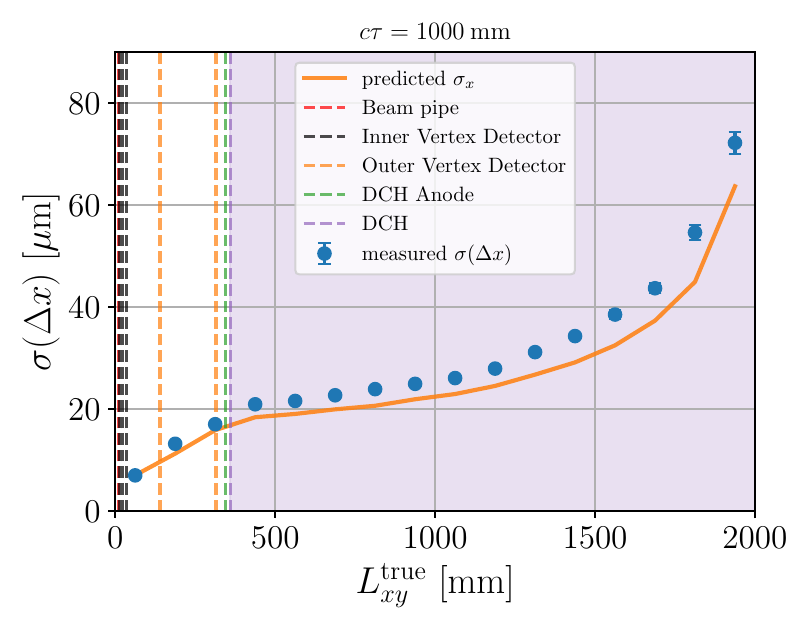}
    \caption{Spatial vertexing resolution in the transverse direction (see text for details), exemplified by $x$, depending on the $N$ decay position $L_{xy}$.
    Data points with (binomial) error bars indicate the bootstrapped residual widths in each bin, while the orange line denotes the median of the post-fit vertex uncertainty.
    Vertical dashed lines denote the different barrel layers of the tracking system as indicated in the legends.}
    \label{fig:resolution_x_Lxy}
\end{figure}

\begin{figure}
    \centering
    \includegraphics[width=0.48\linewidth]{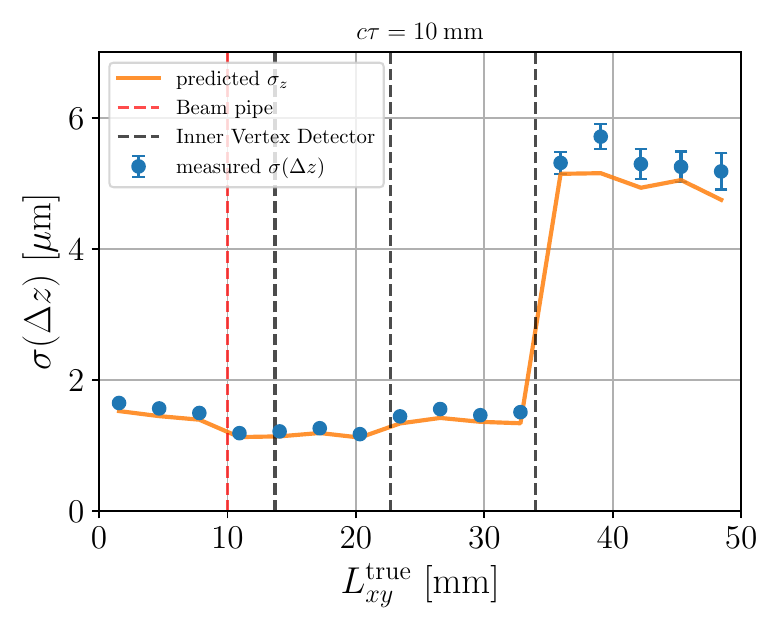}
    \includegraphics[width=0.48\linewidth]{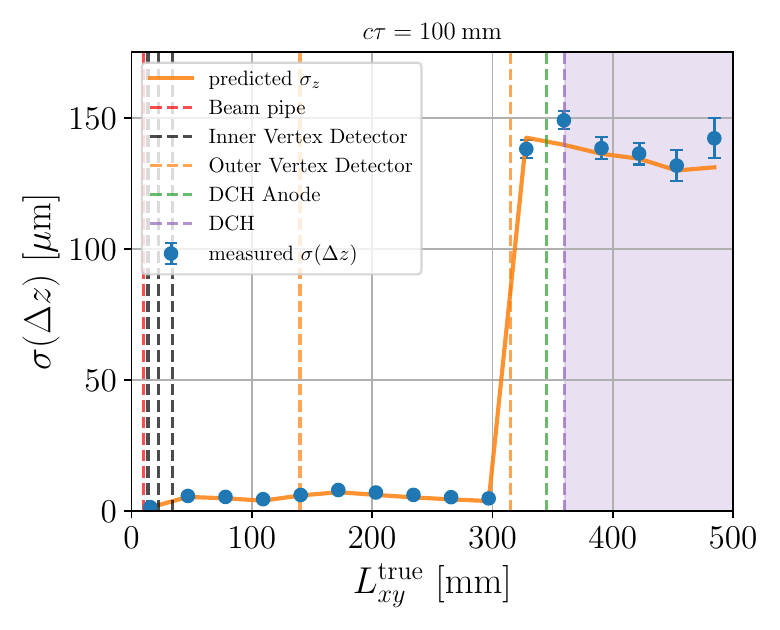}
    \includegraphics[width=0.48\linewidth]{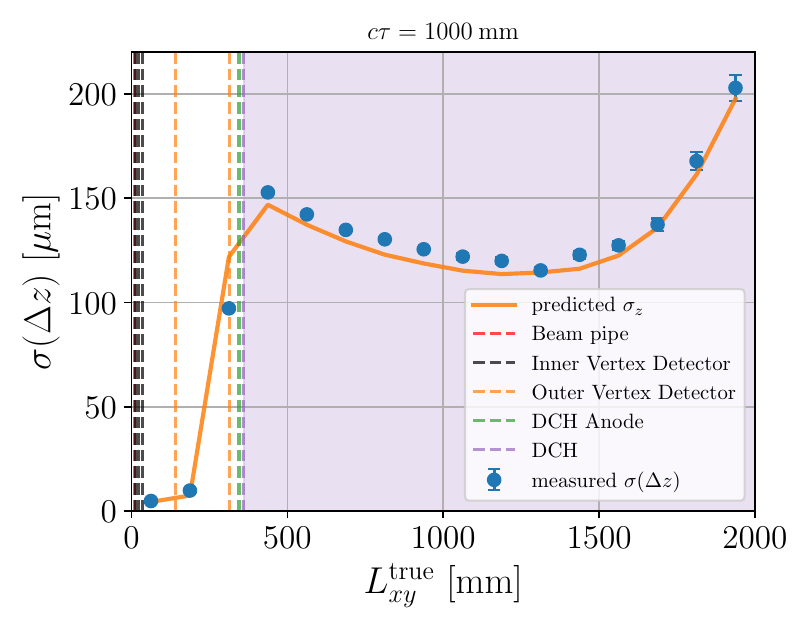}
    \caption{Spatial vertexing resolution in the longitudinal direction (see text for details), depending on the $N$ decay position $L_{xy}$.
    Data points with (binomial) error bars indicate the bootstrapped residual widths in each bin, while the orange line denotes the median of the post-fit vertex uncertainty.
    Vertical dashed lines denote the different barrel layers as indicated in the legends.}
    \label{fig:resolution_z_Lxy}
\end{figure}

The final quality assessment conducted here concerns the vertex resolution.
We match reconstructed vertices to the closest ``true vertex'' (with the purities as outlined in the above) and compute the residuals $\vec v_\mathrm{reco} - \vec v_\mathrm{true}$.
For each bin in $L_{xy}$, the distributions have long non-gaussian tails due to many other observables (e.g. $p_T$, $\eta$, $L_z$) having been marginalised, such that we report the central $68\%$ width of bootstrap resampled distributions in each bin as the ``measured resolution'' and the median of the $\sqrt{\mathcal C_{\vec v}(xx)}$ and $\sqrt{\mathcal C_{\vec v}(zz)}$ distributions as the ``predicted'' transverse and longitudinal uncertainties\footnote{Equivalent results have been found for the $y$-direction but are omitted here for brevity.}.
The results are shown in Figures~\ref{fig:resolution_x_Lxy} and~\ref{fig:resolution_z_Lxy} together with the positions of the different (barrel) layers of the tracking detector.
Owed to a single hit resolution of $3\:\mu$m in the inner  and $7\:\mu$m in the outer vertex detector layers~\cite{2502.21223}, the spatial vertex resolution for decays inside the beam pipe or the inner vertex detector is at $2\:\mu$m or below in both the transverse (we take $x$ as an example with very similar results for $y$) and longitudinal directions.
For decays outside of the inner vertex barrels but inside the outer barrels, the resolution diminishes to $5-12\:\mu$m in the transverse direction, while it stays at around $5\:\mu$m in the longitudinal direction.

The biggest increase in vertex resolution is however for decay vertices in the drift-chamber (DCH), which has a transverse single hit resolution of about $100\:\mu$m, while the $z$-coordinate is obtained from the small stereo angle of the wire planes and is therefore significantly less precise.
Consequently, the vertex resolution in the longitudinal direction ``jumps'' to $100-200\:\mu$m while in the transverse direction it varies between $20-70\:\mu$m.

We further notice that the ``predicted'' and ``measured'' resolutions agree very well with each other and further (binned) pull studies let us estimate that the post-fit vertex covariance slightly underestimates the uncertainties by at most $2-5\%$ in highly populated bins while it increases up to $10\%$ in bins with a sizeable statistical uncertainty.
Finally, the standardised post-fit $\chi^2/\mathrm{ndf}$ distribution of all vertices has a median of $0.997$ with a width of $0.397$. 
Together with the good agreement between ``predicted'' and ``measured'' uncertainties, this validates our fitter and error model and showcases the excellent vertexing performance of both our algorithm as well as the IDEA detector concept.
\section{Long-lived Particle Reconstruction}
\label{sec:LLP}
In addition to reconstructing displaced vertices, we are interested in reconstructing the kinematics of a long-lived particle as accurately as possible.
For leptonic decays without neutrinos in the final state, it is sufficient to sum the reconstructed four-momenta of the final state charged leptons to obtain a proxy for the LLP momentum.
If additional neutral (or missing) energy is present, in the form of neutrinos or photons, the momentum reconstructed from the visible charged particles underestimates the magnitude of the LLP four-momentum and furthermore leads to a different direction. 
Thus, the direction of the reconstructed momentum and the LLP flight direction which we take as
\begin{equation}
    \vec L = \vec v_\mathrm{DV} - \vec v_\mathrm{PV}\,,
\end{equation}
are not aligned.
A useful measure is given by the angle between the two taken as
\begin{equation}
    \cos\theta_\mathrm{DV} = \frac{\vec L\cdot \vec p_\mathrm{visible}}{|\vec L|\cdot|\vec p_\mathrm{visible}|}\,.
\end{equation}
Even if the LLP momentum has been perfectly reconstructed, $\cos\theta_\mathrm{DV}$ can still significantly differ from $1$ which can then hint at a non-prompt production process of the LLP, which is for instance the case in displaced chain decays.

Without a constrained kinematic fit that leverages information of the production process (if available), we can still find a good proxy on the invariant mass of the decaying LLP in the form of the so-called ``corrected mass''.
Under the approximation that additional invisible energy is aligned with $\vec L$, we can take the transverse momentum
\begin{equation}
    p_\perp = \frac{|\vec p_\mathrm{visible}\times \vec L|}{|\vec L|}\,,
\end{equation}
and define the ``corrected mass'' as
\begin{equation}
    m_\mathrm{corr} = \sqrt{m^2_\mathrm{visible} + p_\perp^2} + p_\perp\,,
\end{equation}
which can be interpreted as a lower bound on the true invariant mass.
If the production mode is known, a kinematic fit can be conducted that further leverages the fitted vertex time with the methods described in~\cite{Aleksan:2024hyq}.
Here, we stay agnostic of the production mode and leave further improvements of the kinematical reconstruction for future work.

For semi-leptonic and hadronic LLP decays, a significant amount of neutral energy is present in the form of photons and neutral hadrons. 
While not directly visible in the tracker, it is deposited in the calorimeters and clustered into jets.
In order to associate the neutral energy in form of calorimeter towers or neutral particle flow candidates with the displaced vertices, we introduce a momentum-weighted jet matching procedure.
For each vertex and for each jet we define an overlap score as
\begin{equation}
    s_\mathrm{overlap} = \frac{2 \sum_{i\in \mathrm{shared}} (p_T^i)^\alpha}{\sum_{i\in\mathrm{jet}} (p_T^i)^\alpha + \sum_{i\in\mathrm{DV}}(p_T^i)^\alpha}\,,
\end{equation}
in which $i\in$~shared counts the tracks associated to both the DV and the jet and $\alpha$ is a tunable parameter such that $\alpha = 1$ leads to a $p_T$ weighted overlap fraction while $\alpha = 0$ leads to an unweighted overlap fraction~\footnote{In our numerical studies both choices for $\alpha$ lead to comparable results, but miss-associations happened more rarely using $\alpha = 1$.
The effect of other values of $\alpha$ have not yet been studied and are left for future work.}.
We then greedily match jets to vertices with the highest overlap score if it exceeds a threshold value of e.g. $s_\mathrm{overlap}\geq 0.1$.

With jets (and potentially isolated leptons) matched to displaced vertices, we can now take
\begin{equation}
    p_\mathrm{LLP}^\mu = \sum_{i\in\text{jets, leptons}} p_i^\mu
\end{equation}
as a proxy for the reconstructed LLP four-momentum.
This jet matching and momentum reconstruction scheme has been implemented as a public \textsc{Delphes} module dubbed \textsc{LLPReconstruction} and is available on GitHub at~\cite{LLP}.

Beyond such a simple matching, we could also associate neutral energy in the form of calorimeter towers or neutral particle flow candidates as ``neutral tracks'' to the displaced vertices.
Such neutral particle flow candidates have typically a much worse spatial resolution than charged tracks (or particle flow candidates), such that they usually do not enter the vertex fit directly and one could use the pointing information of the reconstructed neutral momenta from the calorimeter surface to the DV for an alternative matching scheme.
This is however beyond the scope of the present work and is left for future studies.

With the simple jet matching schemed outlined in the above, and for the same samples used in Section~\ref{sec:validation_and_performance}, after vertex fitting (and track re-fitting), we use \textsc{FastJet-3.5.1}~\cite{Cacciari:2011ma} for jet clustering.
Before charged particle flow candidates are passed to \textsc{FastJet}, we remove isolated leptons (electrons and muons) from the collection~\cite{Amjad:2013tlv,1404.4294}.
We take (very) loose lepton isolation criteria and require a minimum transverse momentum of $p_T^\text{min}\geq 1.0\:\mathrm{GeV}$ and allow for up to $50\%$ additional momentum within an angular cone of $\Delta R\leq 0.1$ in the vicinity of the lepton.
The remaining charged and neutral particle flow candidates are then passed to \textsc{FastJet} and we use the exclusive Durham $k_T$ algorithm~\cite{Catani:1991hj}, reconstructing always exactly 4 jets~\footnote{For the very clean event topologies in this study we did not observe any significant differences between the Durham $k_T$ and the more modern Valencia algorithm~\cite{1404.4294} that allows for more flexible distance measures. The Valencia algorithm could however yield significant improvements in very busy events with many jets.}.

\begin{figure}
    \centering
    \includegraphics[width=0.48\linewidth]{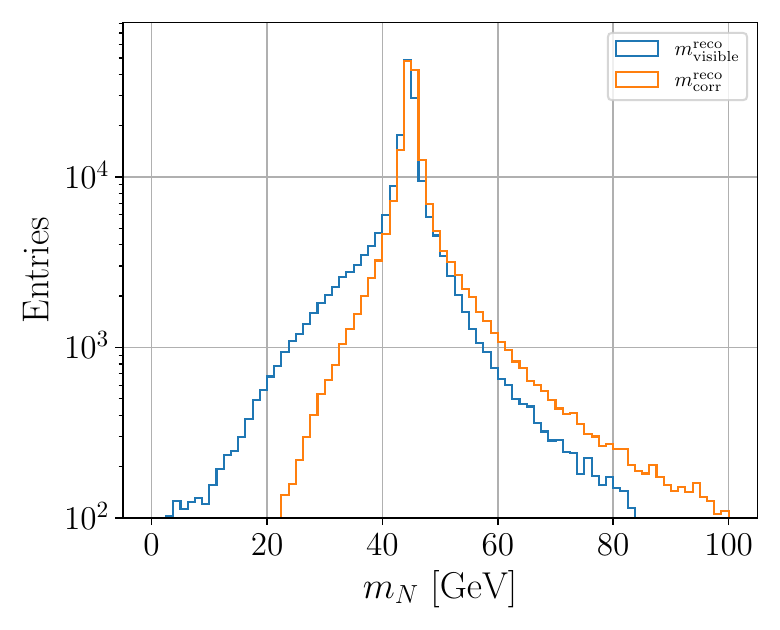}
    \includegraphics[width=0.48\linewidth]{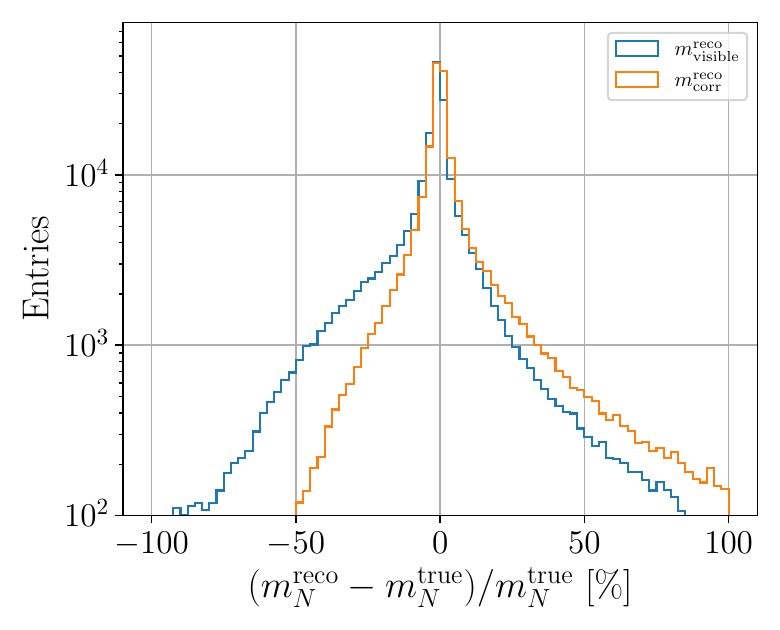}
    \caption{Reconstructed invariant mass distributions for $m_\mathrm{visible}$ (blue) and $m_\mathrm{corr}$ (orange).
    On the left are the absolute distributions, while on the right we present the relative resolution for $m_N^\mathrm{true} = 45\:\mathrm{GeV}$.}
    \label{fig:inv_mass}
\end{figure}

Relying on the sample with $c\tau(N) = 100\:\mathrm{mm}$, we select events in which two displaced vertices are reconstructed and match them to jets and isolated leptons based on the overlap score outlined in the above.
In Figure~\ref{fig:inv_mass} we present the reconstructed invariant mass distribution of the HNLs, both for the ``raw'' visible momentum (summed jet and lepton momenta) as well as the corrected mass $m_\mathrm{corr}$, together with the relative mass resolution.
As one can see, a large fraction of the events are contained within a a window of $m_\mathrm{visible}^\mathrm{reco}\in[0.5 \:m_N,\, 1.2\: m_N]$ or $m_\mathrm{corr}^\mathrm{reco}\in[0.8 \:m_N,\, 1.5\: m_N]$, such that one can place efficient cuts on either $m_\mathrm{visible}$ or $m_\mathrm{corr}$ to reject backgrounds while retaining the majority of the signal events. 

\begin{figure}
    \centering
    \includegraphics[width=0.48\linewidth]{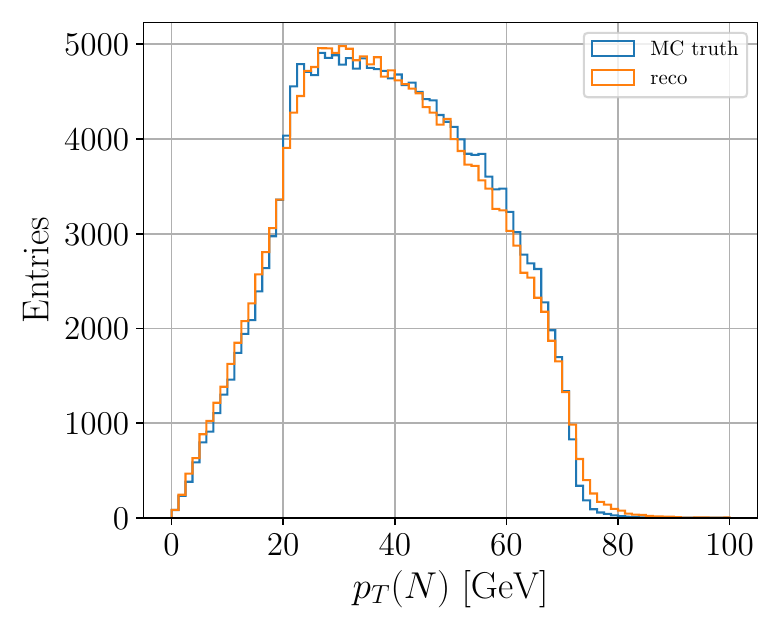}
    \includegraphics[width=0.48\linewidth]{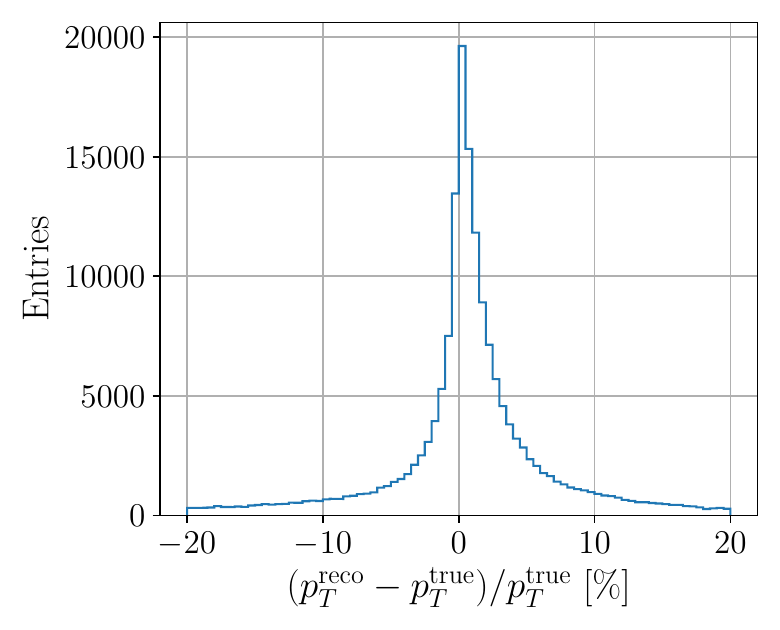}
    \caption{Distribution of the reconstructed transverse momentum $p_T$ of the HNL together with the true distribution, as well as the relative residuals of the $p_T$ distribution.}
    \label{fig:HNL_pT}
\end{figure}

\begin{figure}
    \centering
    \includegraphics[width=0.48\linewidth]{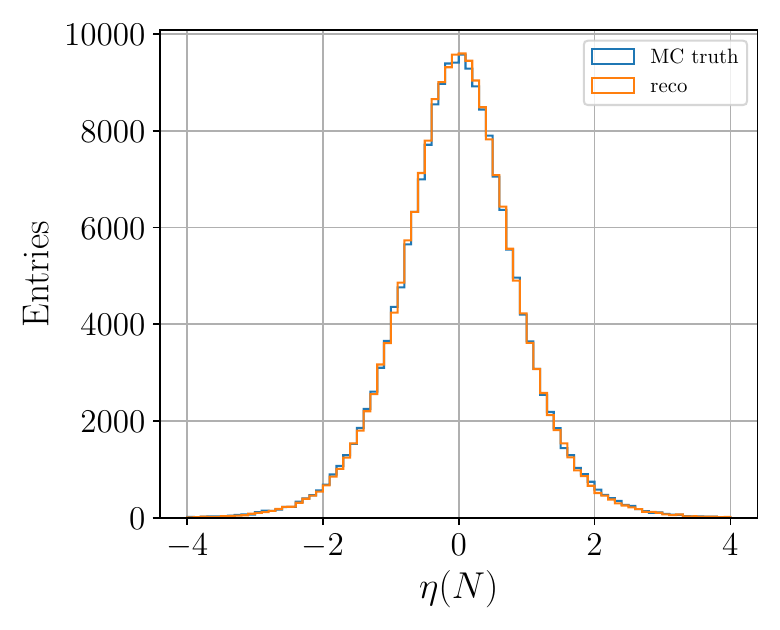}
    \includegraphics[width=0.48\linewidth]{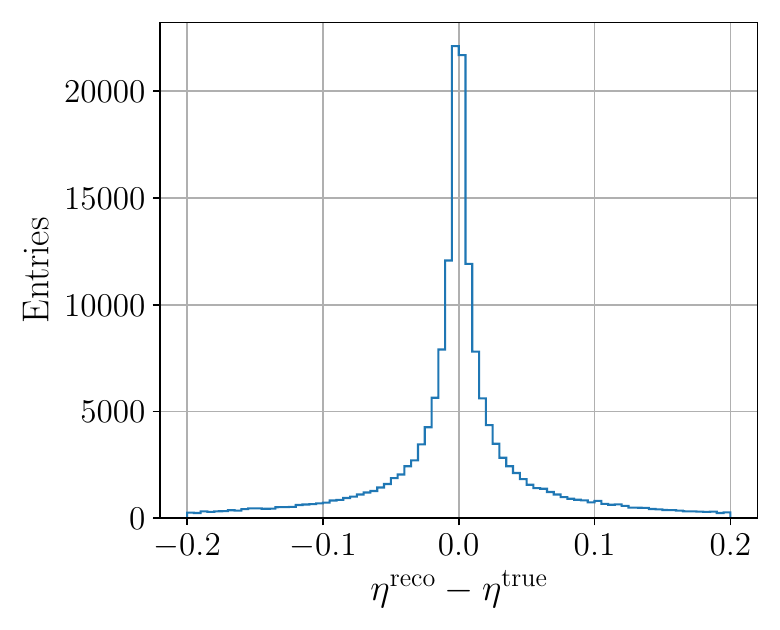}
    \caption{Distribution of the reconstructed pseudo-rapidity $\eta$ of the HNL together with the true distribution, as well as the residuals of the $\eta$ distribution.}
    \label{fig:HNL_eta}
\end{figure}

In addition to the invariant mass of a single HNL, we further show the reconstructed transverse momentum ($p_T$) and pseudo-rapidity ($\eta$) distributions in Figures~\ref{fig:HNL_pT} and~\ref{fig:HNL_eta} together with the Monte-Carlo truth distributions and their (relative) residuals.
The results show an excellent agreement between the reconstructed and true kinematics of the HNL, which encourages us to reconstruct the ``parent'' kinematics, in this case the four-momentum of the decaying Higgs boson, as well.

\begin{figure}
    \centering
    \includegraphics[width=0.48\linewidth]{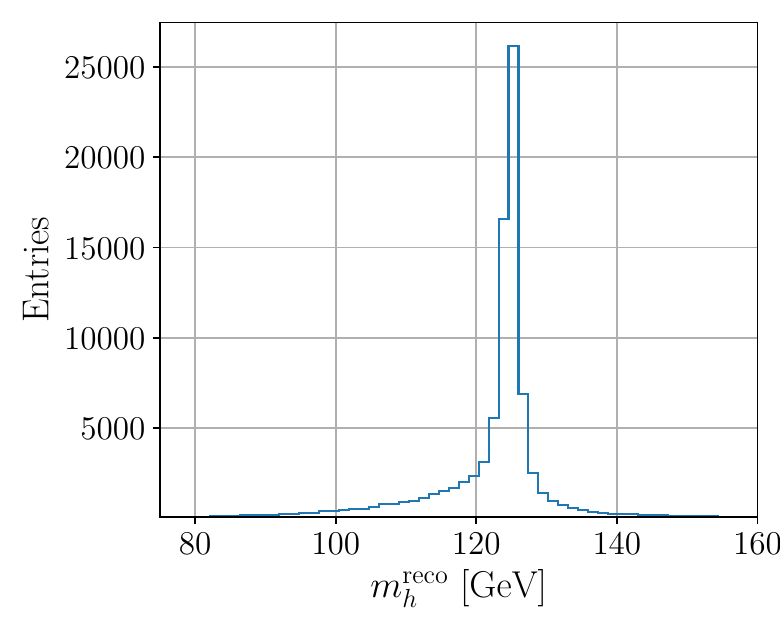}
    \includegraphics[width=0.48\linewidth]{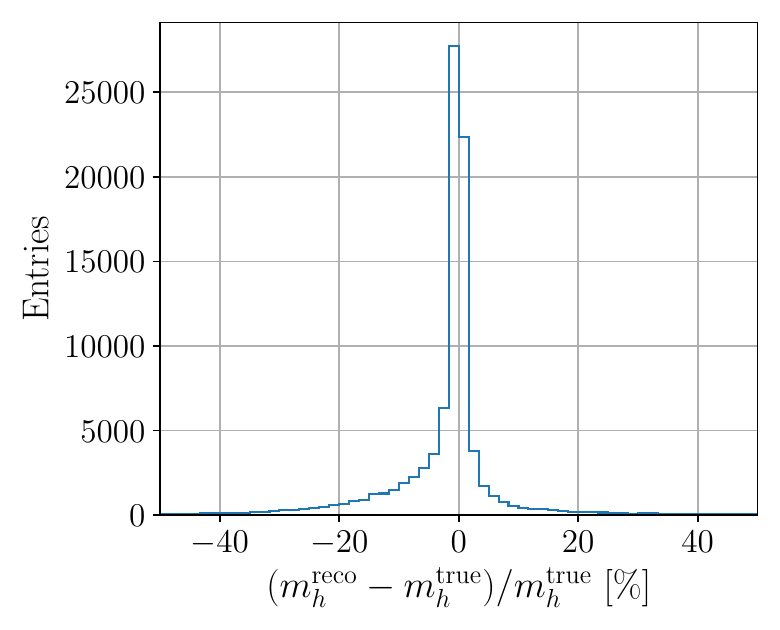}
    \caption{Reconstructed invariant mass distribution of the ``parent particle'' of the two reconstructed HNLs (left), together with the relative residuals (right).}
    \label{fig:inv_mass_h}
\end{figure}

\begin{figure}
    \centering
    \includegraphics[width=0.48\linewidth]{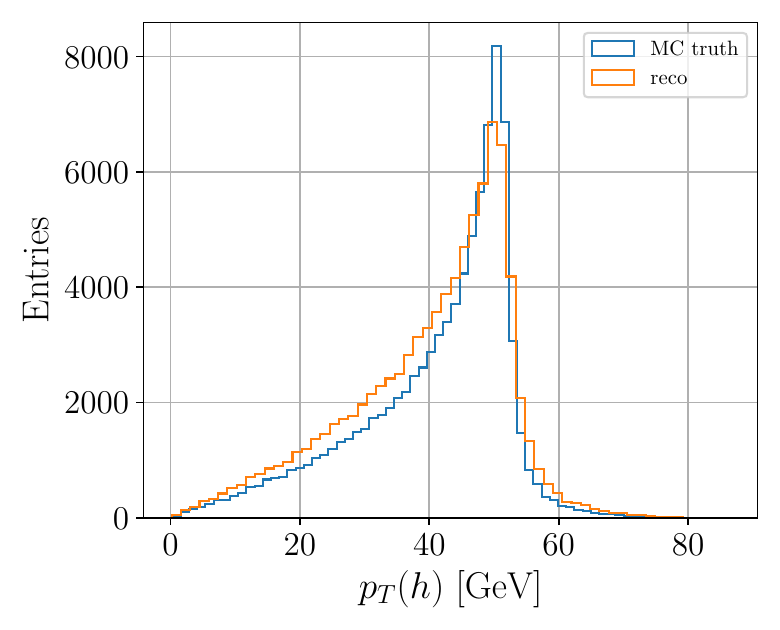}
    \includegraphics[width=0.48\linewidth]{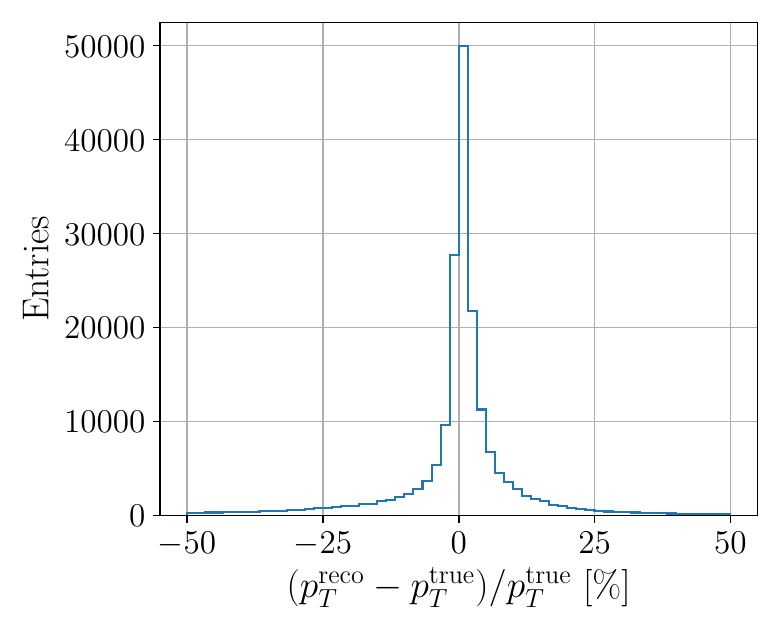}
    \caption{Distribution of the reconstructed transverse momentum $p_T$ of the Higgs together with the true distribution, as well as the relative residuals of the $p_T$ distribution.}
    \label{fig:h_pT}
\end{figure}

\begin{figure}
    \centering
    \includegraphics[width=0.48\linewidth]{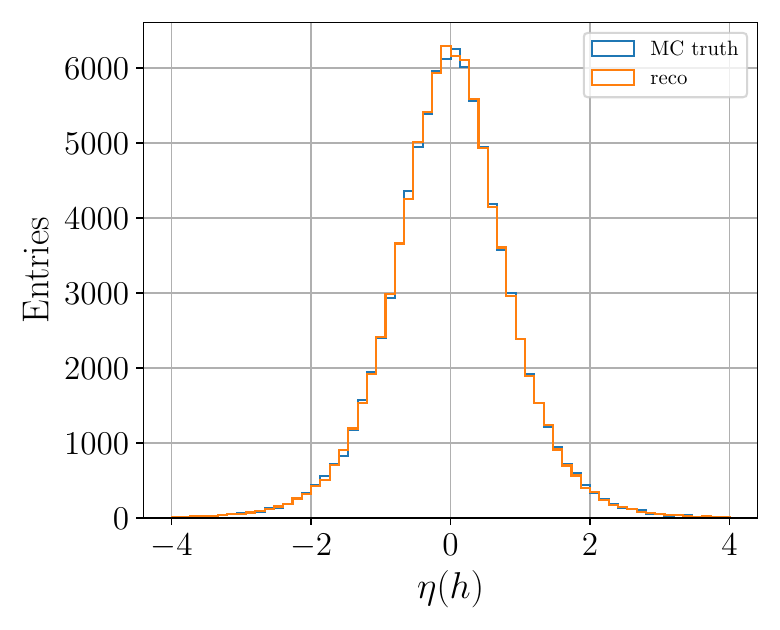}
    \includegraphics[width=0.48\linewidth]{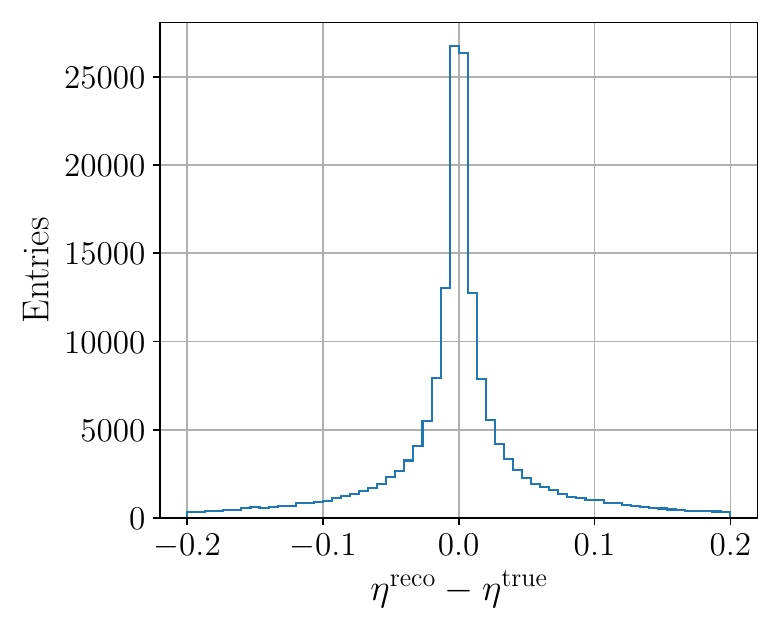}
    \caption{Distribution of the reconstructed pseudo-rapidity $\eta$ of the Higgs together with the true distribution, as well as the residuals of the $\eta$ distribution.}
    \label{fig:h_eta}
\end{figure}

Selecting events with exactly two reconstructed HNLs, we take the Higgs momentum as
\begin{equation}
    p_h^\mu = p_{N_1}^\mu + p_{N_2}^\mu\,.
\end{equation}
As can be seen in Figure~\ref{fig:inv_mass_h}, the reconstructed mass invariant distribution is strongly peaked around the true mass and the vast majority of the events is contained around $m_h\pm 20\%$.
Furthermore, the reconstructed transverse momentum and pseudo-rapidity distributions in Figures~\ref{fig:h_pT} and~\ref{fig:h_eta}, shows a remarkable agreement between the reconstructed and true distributions.

This remarkable agreement between the reconstructed and true kinematics is also owed to the clean nature of the event topology as well as of the simulated detector.
In events with more hadronic activity we expect the performance to somewhat degrade, since mis-associations of neutral particle flow candidates to ``wrong'' jets as well as mis-associations of jets to the ``wrong'' vertex will become more common and would require dedicated mitigation strategies.
Nevertheless, the presented results are very encouraging and form the basis for the example analysis in the following section.
\section{Projected FCC-ee sensitivity on $\mathrm{BR}(h\to NN)$}
\label{sec:analysis}
Based on the kinematic LLP reconstruction introduced in~\ref{sec:LLP}, we perform a full analysis of the potential FCC-ee sensitivity and derive model independent projections for upper bounds on $\mathrm{BR}(h\to NN)$ that could be established with the full dataset to be taken at $\sqrt{s} = 240\:\mathrm{GeV}$.
Assuming the benchmark integrated luminosity of~\cite{2505.00272}, about $\simeq 2.2\times 10^{6}$ $Zh$ events are expected.

We simulate $e^+ e^-\to Z h, h\to NN$ with $N\to\mu^\pm j j$ and $Z\to e^+e^-\,,\mu^+\mu^-, \nu\bar \nu$ over a wide range of masses $m_N$ and proper lifetimes $c\tau_N$ with $m_N\in [5, 10, \dots,60]$~GeV and $c\tau_N\in[0.1, 0.3, 0.5, 1, 3, \dots10000]$~mm with $10^4$ events per mass/lifetime point.
We select events with 2 reconstructed HNLs and at most 2 prompt tracks attached to the primary vertex which are required to be leptonic.
The primary vertex is reconstructed from tracks that satisfy $\mathcal S_\mathrm{IP}\leq 3$ (see Eq.~\eqref{eqn:IPsig}) and have a minimum transverse momentum of $p_T\geq 100\:\mathrm{MeV}$.

For the displaced vertex reconstruction we require that reconstructed vertices have a minimum of 3 associated tracks out of which one has to be a muon.
Vertices with a transverse displacement $L_{xy}\leq 0.5\:\mathrm{mm}$ are rejected.
These three requirements heavily suppress backgrounds and fakes from $K_S$ and $\Lambda_0$ ($V_0$-decays) as well as photon conversions which are overwhelmingly two-prong.
While we did not simulate beam-induced background (BIB), we expect that the requirement of at least $3$ tracks per reconstructed vertex will suppress BIB ($\gamma\gamma\to\ell^+\ell^-$ and $\gamma\gamma\to h^+h^-$ pointing towards the beam-pipe) to negligible levels since possible fake vertices from BIB are overwhelmingly two-prong and forward pointing with tiny transverse displacements $L_{xy}$. 
Heavy-flavour decays of $D$ and $B$-mesons as well as $\tau$-leptons are either already suppressed by the vertex requirements but get even more suppressed by the kinematical requirements outlined in the following.

The reconstructed corrected mass of the HNL candidate is selected in the window $m_\mathrm{corr}\in [0.8 \:m_N,1.5\:m_N]$ and for events that satisfy these criteria we further demand the reconstructed Higgs mass (reconstructed from the sum of HNL four-momenta) to lie in $m_h^\mathrm{reco}\in[100\:,150]$~GeV.

In order to estimate backgrounds, we simulated $10^6$ $e^+e^-\to Zh$ and $e^+e^-\to ZZ$ events with one $Z$ decaying leptonically, as well as $10^6$ $e^+e^-\to W^+ W^-$ events in the SM out of which none survived the aforementioned cuts and we thus consider the analysis to be background free~\footnote{Remaining backgrounds can stem from material interactions in the tracking volume or cosmic rays. For both backgrounds dedicated vetos can be employed which should efficiently identify and remove them. These considerations are however beyond the scope of this work.}.
These findings are in line with background studies conducted in~\cite{Ripellino:2024iem,Bellagamba:2025xpd}.

\begin{figure}
    \centering
    \includegraphics[width=0.48\linewidth]{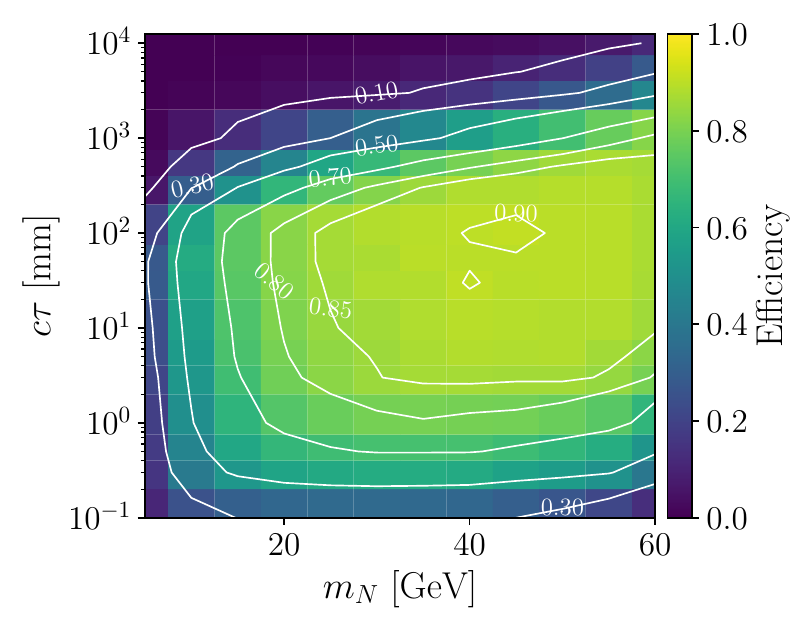}
    \includegraphics[width=0.48\linewidth]{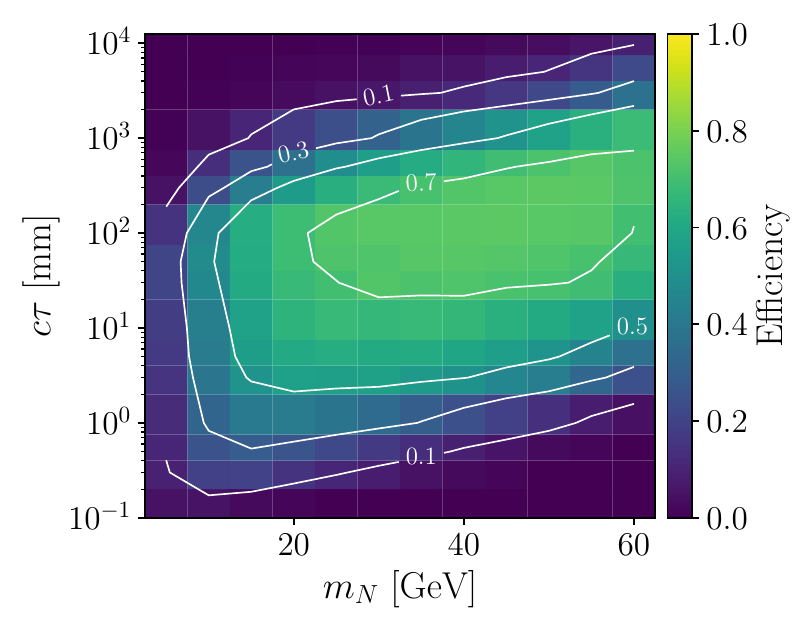}
    \caption{Left: Reconstruction efficiency (including geometrical acceptance) of events with 2 reconstructed displaced vertices. Right: Total selection efficiencies after the additional kinematic selections.}
    \label{fig:signal_efficiencies}
\end{figure}

While the aforementioned cuts allow to reduce backgrounds to negligible levels, signal efficiencies remain very high and are mostly limited by the geometrical acceptance.
In Figure~\ref{fig:signal_efficiencies} we display the total reconstruction efficiencies after vertex reconstruction and demanding that $2$ displaced vertices are reconstructed and the total selection efficiencies after the kinematic/displacement selections outlined in the above.
The reconstruction efficiencies already include the detector acceptance and reach as high as $90\%$, while they sharply drop for very light and very boosted HNL.
The reconstruction efficiencies are consistent with~\cite{Bellagamba:2025xpd}, in which the authors study prompt and displaced signatures of single HNLs from $Z\to N\nu$ decays, relying on the vertex fitting code developed in~\cite{Bedeschi:2024uaf}.
Here, the requirement of at least $3$ tracks attached to the reconstructed vertex further diminishes the reconstruction efficiencies.
The kinematic selection reduce the total efficiencies down to a level of at most $75\%$, also cutting away signal events with a displacement of $L_{xy}\leq 0.5$~mm.
Furthermore, since we only simulated $10^4$ events for each point, we set efficiencies below $1\%$ to 0, as the statistical uncertainty becomes significant in such bins.

\begin{figure}
    \centering
    \includegraphics[width=0.48\linewidth]{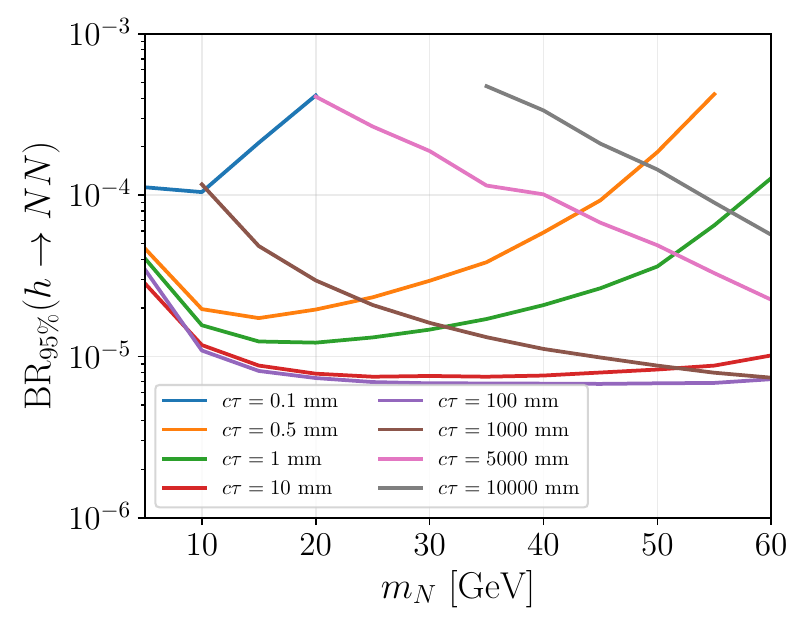}
    \includegraphics[width=0.48\linewidth]{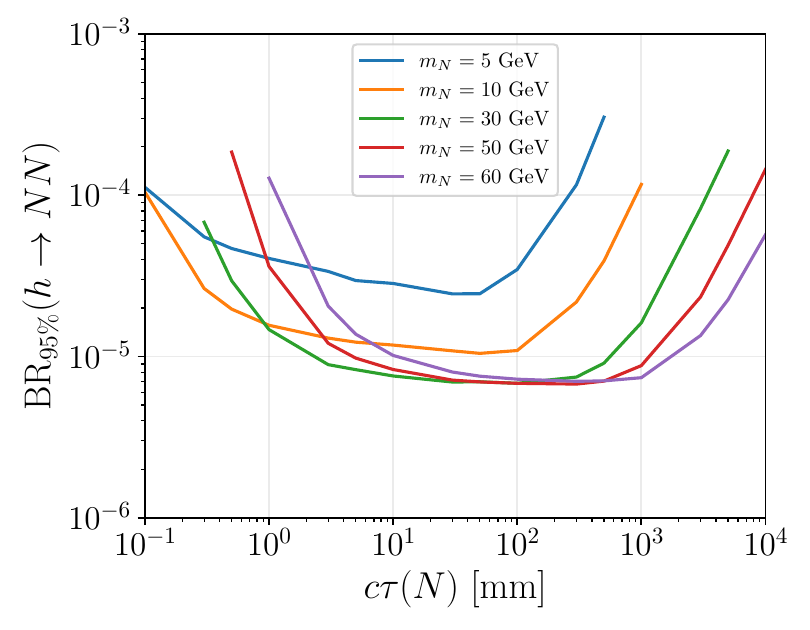}
    \caption{Projected reach on upper limits on $\mathrm{BR}(h\to NN)$ at the $95\%$ confidence level.}
    \label{fig:BRlimits_1D}
\end{figure}

We take as the exclusion of the signal $S$ the approximate significance $\mathcal S \simeq \sqrt{S}$ such that less than $3$ reconstructed events corresponds to an exclusion at the $95\%$-level.
In Figure~\ref{fig:BRlimits_1D} we present the projected reach on $\mathrm{BR}(h\to NN)$ depending on the HNL mass $m_N$ for several choices of the lifetime $c\tau$, as well as depending on the lifetime for several choices of the mass $m_N$.
As can be seen, the projected limits reach down to $\mathrm{BR}_{95\%}(h\to NN)\lesssim (6-7)\times10^{-6}$ for proper lifetimes $10-1000$~mm and masses $20-60$~GeV.
These results are consistent with~\cite{Ripellino:2024iem}, in which the Authors studied exotic Higgs decays to a long-lived scalar $s$, which subsequently decays to $b$-jets.

\begin{figure}
    \centering
    \includegraphics[width=0.65\linewidth]{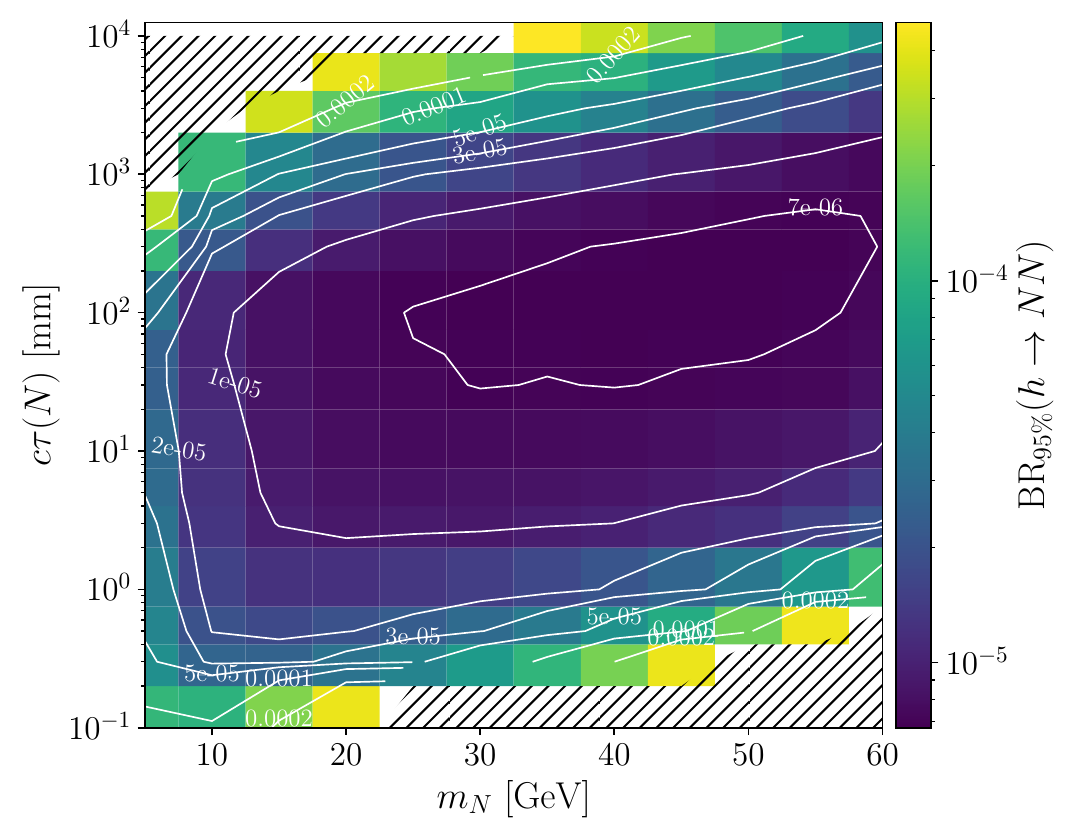}
    \caption{Contours of the projected upper limits on $\mathrm{BR}(h\to NN)$ in the plane of proper lifetime $c\tau$ and HNL mass $m_N$. Hatched areas denote regions in which the efficiency is below $1\%$ and has been set to 0.}
    \label{fig:BRlimits_2D}
\end{figure}

A combined overview can be seen in Figure~\ref{fig:BRlimits_2D}, where we show $2D$-contours of the projected $95\%$ upper limit on $\mathrm{BR}(h\to NN)$.
As can be seen, a wide range of masses and proper lifetimes can be covered and potentially excluded down to a level of $\mathrm{BR}_{95\%}(h\to NN)\lesssim10^{-5}$.
As is the case of the LRSM, these bounds could indirectly constrain the mixing of the SM Higgs sector with a BSM scalar coupled to HNLs.
Within the LRSM, the lifetime of the HNLs is further directly proportional to the breaking scale of the $SU(2)_R$ gauge group via the mass of the additional heavy gauge bosons.
Therefore, bounds on $\mathrm{BR}(h\to NN)$, with long-lived/displaced HNLs could further indirectly constrain the mass-scale of such mediators (see e.g.~\cite{Fuks:2026hra} or ~\cite{Maiezza:2015lza} for an equivalent LHC study).
In order to make these bounds model-independent, one also needs to account for the fact that we assumed here that the HNLs decay semi-leptonically, which is the dominant decay mode in the LRSM for masses $m_N \lesssim m_W$, but might not be the case in other models.
The projected upper bounds should therefore be interpreted as $\mathrm{BR}_{95\%}(h\to NN)\lesssim 10^{-5} \times(\mathrm{BR}(N\to\ell^\pm j j))^{-2}$.

\section{Conclusions}
\label{sec:concs}
In this work we have presented a new approach to displaced vertex reconstruction that combines a graph-based track clustering strategy for vertex finding with a robust vertex fitter. 
The algorithm has been implemented as a self-contained \textsc{Delphes} module, making it a turn-key tool for phenomenological studies of long-lived particles. 
Unlike existing public software frameworks such as \textsc{CMSSW}~\cite{CMSSW}, \textsc{ACTS}~\cite{Ai:2021ghi}, or \textsc{RAVE}~\cite{rave,Waltenberger:2011zz}, which are either experiment-specific or too heavy-weight for fast simulations (since they are designed for full offline reconstruction), our implementation can be straightforwardly integrated into any \textsc{Delphes} detector card and directly used in the standard \textsc{MadGraph}–\textsc{Pythia}–\textsc{Delphes} pipeline.
The code is publicly available on GitHub at~\href{https://github.com/jkriewald/delphes-LLP/}{https://github.com/jkriewald/delphes-LLP/}.

An additional outcome of this work is a brief review of the vertex fitting problem. 
Starting from a unified Gauss-Newton framework, we derive both equivalent Schur-complement formulations, recovering the established transverse-information vertex fit while making explicit its underlying geometric interpretation. 
In particular, profiling the common vertex reveals that vertex fitting can equivalently be understood as the coupled minimisation of pairwise track distances. 
We hope that this presentation makes the underlying principles of modern vertex fitting more accessible and provides a natural framework for future extensions, such as the inclusion of timing information presented here.

We validated the vertex finding algorithm in an IDEA-like FCC-ee detector setup, using the exotic Higgs decay $h\to NN$ as a benchmark process. 
This channel provides a broad coverage of displaced vertex topologies, ranging from millimetre to metre scales, and is motivated by several classes of neutrino mass models such as the Left–Right Symmetric Model~\cite{Maiezza:2015lza,Nemevsek:2016enw,Urquia-Calderon:2023dkf,Fuks:2025jrn,Liu:2025ldf} and extended type-I seesaw scenarios~\cite{2202.07310,Yang:2025jxc}. 
We demonstrated that our method achieves excellent reconstruction efficiency, purity, and resolution across the entire fiducial tracking volume. 
Exploiting the reconstructed vertices for kinematic analyses, we further derived projections for the FCC-ee sensitivity to exotic Higgs branching ratios, reaching the level of $\mathrm{BR}(h\to NN)\lesssim 10^{-5}$ in a wide region of parameter space.

The results presented here highlight both the physics potential of graph-based vertexing for LLP reconstruction and its practical usability for the community. 
Beyond the FCC-ee, our approach is readily applicable to other (future) collider scenarios, including the CEPC, CLIC, or a high-energy muon collider, as well as to hadron collider settings. 
We hope that this turn-key \textsc{Delphes} implementation will help bridge the gap between phenomenological studies and experimental analyses, and provide a flexible basis for more realistic LLP phenomenology and sensitivity projections in the future.

\section*{Acknowledgements}
This project is supported by the Slovenian Research Agency under the research core funding 
No. P1-0035 and in part by the research grants J1-3013 and N1-0253, and has received further support by the bilateral project Proteus PR-12696.
I thank Fabrizio Nesti for many useful discussions and Miha Nemevšek for careful reading of the manuscript.
I am further grateful for the kind hospitality of LPTHE where this work has been initiated.

\appendix
\section{Track Model}
\label{sec:trackmodel}
While the derivation of the Vertex fit is independent of the track model, for concreteness we outline here the track model used in the numerical studies.
Specifically, we work in the perigee parametrisation~\cite{Billoir:1992yq}, that is a local helix parametrisation in which a track state can be expressed via five parameters 
\begin{equation}
    \vec\alpha = (D_0, \varphi_0, C, z_0, \cot\theta)^T\,,
\end{equation} 
evaluated at a reference surface, e.g. the track's closest approach to the beam-line.
The track parameters are $D_0$ the signed transverse impact parameter, $\varphi_0$ the azimuthal angle, $z_0$ the longitudinal impact parameter and $\theta$ the polar angle.
Finally $C$ is the track curvature given by
\begin{equation}
    C = - \frac{q B_z c}{2 p_T}\,,
\end{equation}
where $q$ is the particle's charge, $B_z$ the longitudinal magnetic field, $c$ the speed of light and $p_T$ the transverse momentum.
Since we assume a constant magnetic field along the $z$-direction, we keep the perigee to the beam-line as a fixed reference surface~\footnote{Beyond this idealisation, the track's reference surface should be updated at each re-linearisation point by solving the particles equation of motion in a non-homogeneous magnetic field and one further needs to account for interaction with the detector material such as multiple scattering effects, as is for instance done in ACTS~\cite{Ai:2021ghi} via the ``eigenstepper''.}.
The particle position on the helix can then be expressed as
\begin{equation}
    \vec x(\varphi) = \begin{pmatrix}-D_0 \sin\varphi_0 + \frac{1}{2C}(\sin(\varphi + \varphi_0) - \sin\varphi_0)\\
    D_0 \cos\varphi_0 - \frac{1}{2C}(\cos(\varphi + \varphi_0) - \cos\varphi_0)\\
    z_0 + \frac{\varphi\cot\theta}{2C}\end{pmatrix}\,,\label{eqn:trkX}
\end{equation}
in which $\varphi$ is the phase advance on the helix in radians.
For numerical purposes however it is advantageous to work in ``true arc-length'' instead of the azimuthal phases, which gives scales equally for all tracks independent of their transverse momentum $p_T$ (and therefore their curvature).
The arc length, meaning the distance travelled along the helix between two reference phases $\varphi_{1,2}$, is given in an integral form by
\begin{equation}
    s = \int_{\varphi_1}^{\varphi_2}\left|\frac{\partial\vec X(\varphi)}{\partial\varphi}\right| d\varphi = \frac{\sqrt{1 + \cot^2\theta}}{2C}(\varphi_2 - \varphi_1)\,,
\end{equation}
so that for $\varphi_1 = 0$ and $\varphi_2\equiv \varphi$ we get a linear map between the azimuthal phase shift and the helix arc length
\begin{equation}
    \varphi(s) = \frac{2C}{n}s\,,
\end{equation}
with $n = \sqrt{1 + \cot^2\theta}$.
For convenience, we also list the necessary derivatives with respect to $s$,
\begin{equation}
    \vec t(s) \equiv \frac{\partial \vec x}{\partial s} = \frac{1}{n}
    \begin{pmatrix}
        \cos(\varphi_0 +\frac{2 C}{n}s )\\
        \sin(\varphi_0 +\frac{2 C}{n}s )\\
        \cot\theta
    \end{pmatrix}\,,\label{eqn:trkdXds}
\end{equation}
which is conveniently of unit-length $|\vec t(s)| = 1$.
 
The derivative with respect to the track parameters $\vec\alpha$ is given by
\begin{equation}
    J_\alpha(s,\vec\alpha) \equiv \frac{\partial \vec x(s,\vec\alpha)}{\partial\vec\alpha}\,,\label{eqn:trkJx}
\end{equation}
with the columns
\begin{eqnarray}
    \frac{\partial \vec x(s,\vec\alpha)}{\partial D_0} &=& 
    \begin{pmatrix}
        -\sin\varphi_0   \\
        \cos\varphi_0\\0
    \end{pmatrix}\,,\:\: \frac{\partial \vec x(s,\vec\alpha)}{\partial \varphi_0} = 
    \begin{pmatrix}
        -D_0 \cos\varphi_0 + \frac{1}{2C}\left(-\cos\varphi_0 + \cos(\varphi_0 + \frac{2C}{n}s\right)\\
        -D_0 \sin\varphi_0 - \frac{1}{2C}\left(-\sin\varphi_0 + \sin(\varphi_0 + \frac{2C}{n}s\right)\\
        0
    \end{pmatrix}\nonumber\\
    \frac{\partial \vec x(s,\vec\alpha)}{\partial Z_0} &=& 
    \begin{pmatrix}
        0   \\
        0   \\
        1
    \end{pmatrix}\,,\:\:
    \frac{\partial \vec x(s,\vec\alpha)}{\partial C} = 
    \begin{pmatrix}
        \frac{s}{C n}\cos(\varphi_0 + \frac{2C}{n}s) - \frac{1}{2C^2}(-\sin\varphi_0 + \sin(\varphi_0 + \frac{2C}{n}s)\\
        \frac{s}{C n}\sin(\varphi_0 + \frac{2C}{n}s) + \frac{1}{2C^2}(-\cos\varphi_0 + \cos(\varphi_0 + \frac{2C}{n}s)\\
        0
    \end{pmatrix}\,,\nonumber\\
    \frac{\partial \vec x(s,\vec\alpha)}{\partial \cot\theta} &=& 
    -\frac{1}{n^3}\begin{pmatrix}
     s\cot\theta\cos(\varphi_0 + \frac{2C}{n} s)\\
     s\cot\theta\sin(\varphi_0 + \frac{2C}{n} s)\\
     -s
    \end{pmatrix}\,.
\end{eqnarray}
With the help of this Jacobian, we can transport the positional covariance of the measured track parameters under the simplifying assumption of linear Gaussian error propagation as
\begin{equation}
    \mathcal C_x(s,\vec\alpha) = J_x(s,\alpha)\, \mathcal C_\alpha \,J_x(s,\alpha)^T\,,\label{eqn:trkCx}
\end{equation}
in which $\mathcal C_\alpha$ is the measured covariance matrix of the track parameters, established by the track fit.

\section{Usage in \textsc{Delphes}}
\label{sec:delphes}
The Vertex finder and fitter described in Section~\ref{sec:vertex_fitting} and ~\ref{sec:graph} have been implemented as a self-contained \textsc{Delphes} module dubbed \textsc{GraphDisplacedVertexFinder} (available at~\cite{Graph}), while the kinematic reconstruction of LLPs via the jet matching procedure described in Section~\ref{sec:LLP} is in a second \textsc{Delphes} module called \textsc{LLPReconstruction} (available on GitHub at~\cite{LLP}).
In order to parse the new objects a few minor additions to the basic \textsc{Delphes} classes and the root \textsc{TreeWriter} had to be made, so that the entire repository available at \href{https://github.com/jkriewald/delphes-LLP/}{https://github.com/jkriewald/delphes-LLP/} has to be downloaded/cloned but can be compiled and used as usual.
An example card can be found at~\cite{Card}.
Here we describe the basic tunable parameters of both modules in Tables~\ref{tab:GraphDV} and~\ref{tab:LLPReco}.

The \textsc{GraphDisplacedVertexFinder} module takes as its input \textsc{Tracks} or \textsc{EflowTracks} that have a full rank covariance matrix for the track parameters, which can be provided by the \textsc{TrackSmearing} or \textsc{TrackCovariance} modules of standard \textsc{Delphes}.
If timing should be used (and fitted), the \textsc{TimeSmearing} module is further required to have been executed upstream.
As its output, the \textsc{GraphDisplacedVertexFinder} module gives a collection of displaced vertices, primary vertices~\footnote{So far we do not support primary vertex finding under pile-up conditions such that only one primary vertex is searched for and fitted.} and modified tracks (track parameters refitted).
\renewcommand\arraystretch{1.3}
\begin{table}[ht]
    {
    \small
    \centering
    \begin{tabular}{|c|c|p{0.6\textwidth}|}
    \hline
    Parameter name & Default & Description\\
    \hline
      \texttt{MinTrackIPSig}   &  3.0 & Impact parameter significance to identify displaced tracks\\
      \texttt{MinD0} & 0.0 [mm]& Additional cut on minimum transverse track displacement\\
      \texttt{MinZ0} & 0.0 [mm]& Additional cut on minimum longitudinal track displacement\\
      \texttt{MinTrackPT} & 0.1 [GeV] & Minimum track $p_T$ to be considered in the vertex finder\\
      \hline
      \texttt{BeamSpotSigmaX} & 0.01 [mm] & Beam-spot spread in the $x$ direction (used for prior in PV fit)\\
      \texttt{BeamSpotSigmaY} & 0.01 [mm] & Beam-spot spread in the $y$ direction \\
      \texttt{BeamSpotSigmaZ} & 1.0 [mm] & Beam-spot spread in the $z$ direction \\
      \texttt{PVChi2NDFMax} & 9.0 & Maximum $\chi^2/\mathrm{ndf}$ for the PV fit to be acceptable\\
      \texttt{UsePVCut} & \texttt{true} & Switch to enable the PV compatibility to tag displaced tracks rather than ``raw'' impact parameters\\
      \texttt{PVCutChi2} & 9.0 & Maximum $\chi^2$ compatibility to the PV for a track to be considered prompt\\
      \hline
      \texttt{SeedSelector} & -1 & Demand a ``special seed track'' to be present (-1 = no requirement, 0 = larger $p_T$ cut, 1 = large IP significance, 2 = high-$p_T$ lepton)\\
      \texttt{MinSeedPT} & 0.1 [GeV] & Minimum $p_T$ for a given seed track\\
      \texttt{MinSeedIPSig} & 5.0 & Larger IP significance for a given seed track\\
      \texttt{MinTracks} & 3 & Minimum amount of tracks for vertex candidate\\
      \texttt{kNN} & -1 & Mutual $k$-NN pruning (-1 = no pruning)\\
      \texttt{MinSupport} & 1 & Minimum triplet support for a graph edge\\
      \texttt{PruneBridges} & \texttt{true} & Enable bridge pruning\\
      \texttt{BridgeCut} & 3.0 & Stronger $\chi^2$ cut for bridge pruning\\
      \hline
      \texttt{chi2PairCut} & 9.0 & Maximum $\chi^2/\mathrm{ndf}$ for pair fits in the graph building stage\\
      \texttt{chi2TripletCut} & 9.0 & Maximum $\chi^2/\mathrm{ndf}$ for triplet fits in the graph building stage\\
      \texttt{weightCut} & 0.5 & Minimum weight for a track to stay attached to a vertex after the fit has converged\\
      \texttt{Chi2NDFMax} & 9.0 & Maximum $\chi^2/\mathrm{ndf}$ for a vertex fit to be acceptable\\
      \texttt{UseTiming} & \texttt{true} & Use timing compatibility in pair fits at the graph building stage\\
      \texttt{TimeGate} & 9.0 & Maximum timing $\chi^2$ for the timing gate\\
      \texttt{Chi2AssociationMax} & 9.0 & Maximum single track to vertex $\chi^2$ for leftover tracks to be assigned to their best vertex\\
      \hline
      \texttt{chi2\_0} & 9.0 & Sigmoid width\\
      \texttt{beta} & 2.0 & Sigmoid slope\\
    \hline
    \end{tabular}
    \caption{Settings for the \textsc{GraphDisplacedVertexFinder} module.}
    \label{tab:GraphDV}
    }
\end{table}
\renewcommand\arraystretch{1.0}

The \textsc{LLPReconstruction} module takes as its input the displaced vertices found by \textsc{GraphDisplacedVertexFinder} and jets that (crucially) must have been clustered with the modified \textsc{Tracks}/\textsc{EflowTracks} and the neutral \textsc{Eflows}.
The jet clustering must have been run on the \textsc{Tracks}/\textsc{EflowTracks} outputted by \textsc{GraphDisplacedVertexFinder}, because the overlap is computed based on pointer identity.
The output of \textsc{LLPReconstruction} are LLP candidates with reconstructed four-momenta, as well as jets and light leptons that have been tagged as displaced.
\renewcommand\arraystretch{1.3}
\begin{table}[ht]
    {\small
    \centering
    \begin{tabular}{|c|c|p{0.6\textwidth}|}
    \hline
    Parameter name & Default & Description\\
    \hline
    \texttt{cosThetaCut} & 0 & Alignment cut to reject LLP candidates with momenta that do not align wih the flight direction\\
    \texttt{MomentumThreshold} & 0.8 & Reject mis-reconstructed LLP candidates with $|\vec p_\mathrm{jets}| < 0.8|\vec p_\mathrm{tracks}|$\\
    \texttt{MaxDeltaR} & 0.3 & Set tie-breaker alignment score for jet matching $\propto \exp(-(\Delta R(\text{LLP}, \text{jet}))^2/(\Delta R_\text{max})^2)$, only used as a tie-breaker if a jet has identical overlap with two DVs \\
    \texttt{OverlapThreshold} & 0.1 & Minimum $p_T$-weighted track overlap to match a jet to a DV\\
    \texttt{PTAlpha} & 1.0 & Exponent of $p_T$ in the $p_T$-weighted overlap score\\
    \texttt{MinLeptonPT} & 1.0 [GeV] & Minimum lepton $p_T$ to be tagged as displaced (does not affect LLP momentum reconstruction)\\
    \hline
    \end{tabular}
    \caption{Settings for the \textsc{LLPReconstruction} module.}
    \label{tab:LLPReco}
    }
\end{table}
\renewcommand\arraystretch{1.0}
{
\bibliography{bibliography.bib}
\bibliographystyle{JHEP}
}
\end{document}